\documentclass[structabstract]{aa}  
\usepackage{graphicx}
\usepackage{txfonts}
\usepackage{natbib}
\begin{document}

\title{Magnetic field amplification and electron acceleration to near-energy equipartition with ions by a mildly relativistic quasi-parallel plasma protoshock.}

   \author{G. C. Murphy
          \inst{1}
          \and
          M.E. Dieckmann\inst{2}
          \and
          A. Bret\inst{3}
          \and
          L. O'C. Drury\inst{1}
          }

   \institute{Dublin Institute for Advanced Studies, 
              31 Fitzwilliam Place, Dublin 2, Ireland\\
              \email{\{gmurphy,ld\}@cp.dias.ie}
         \and
            Department of Science and Technology, Link\"oping University, SE-60174 Norrk\"oping, Sweden\\
             \email{Mark.E.Dieckmann@itn.liu.se}
         \and
            ETSI Industriales, Universidad de Castilla-La Mancha, 13071 Ciudad Real, Spain\\
             \email{AntoineClaude.Bret@uclm.es}
             }

\titlerunning{Field amplification and electron acceleration in a protoshock}

\date{Received, accepted}

\abstract
{The prompt emissions of gamma-ray bursts (GRBs) are seeded by radiating ultrarelativistic electrons. Kinetic 
energy dominated internal shocks propagating through a jet launched by a stellar implosion, are expected
to dually amplify the magnetic field and accelerate electrons.
}
{We explore the effects of density asymmetry and of a quasi-parallel magnetic field on the collision of two plasma clouds.}
{A two-dimensional relativistic particle-in-cell (PIC) simulation models the collision with 0.9c of two plasma 
clouds, in the presence of a quasi-parallel magnetic field. The cloud density ratio is 10. The densities of ions 
and electrons and the temperature of 131 keV are equal in each cloud, and the mass ratio is 250. The peak Lorentz 
factor of the electrons is determined, along with the orientation and the strength of the magnetic field at the 
cloud collision boundary.}
{The magnetic field component orthogonal to the initial plasma flow direction is amplified to values that exceed 
those expected from the shock compression by over an order of magnitude. The forming shock is quasi-perpendicular 
due to this amplification, caused by a current sheet which develops in response to the differing
deflection of the upstream electrons and ions incident on the magnetised shock transition layer. The 
electron deflection implies a charge separation of the upstream electrons and ions; the resulting electric field 
drags the electrons through the magnetic field, whereupon they acquire a relativistic mass comparable to that of the 
ions. We demonstrate how a magnetic field structure resembling the cross section of a flux tube grows 
self-consistently in the current sheet of the shock transition layer. Plasma filamentation develops behind the 
shock front, as well as signatures of orthogonal magnetic field striping, indicative of the filamentation instability. 
These magnetic fields convect away from the shock boundary and their energy density exceeds by far the thermal 
pressure of the plasma. Localized magnetic bubbles form.}
{
Energy equipartition between the ion, electron and magnetic energy is obtained at the shock transition layer.
The electronic radiation can provide a seed photon population that can be energized by secondary processes (e.g. inverse Compton).
}

\keywords{
gamma rays: bursts --
acceleration of particles -- 
shock waves --
magnetic fields --
ISM: jets and outflows --
methods:numerical}

\maketitle

\section{Introduction}
\subsection{Observations and Context}
Gamma ray bursts (GRBs) are eruptions of electromagnetic radiation at 
cosmological distances. One group of GRBs, those with a long duration, 
is attributed to the implosion of supermassive stars. This is supported 
by observations, where GRBs precede supernovae \citep{Hjorth:2003zl} and
of particularly violent stellar explosions that show some resemblances
with GRBs \citep{Kulkarni:1998lh}. GRBs are thought to be signatures of 
plasma ejection from a forming compact object, such as a neutron star 
or a black hole. 

The fireball model due to \citet{Meszaros:1992xq} and \citet{Rees:1994df} 
assumes that the plasma is ejected in form of a highly relativistic 
collimated jet by extreme supernovae (hypernovae) and
that the jet dynamics is kinematically driven. It has been used to 
explain the anisotropic radiation bursts. In this model, plasma 
clouds collide due to the nonuniform flow speed and density of the jet, 
which is a consequence of the nonstationarity of the jet source. The 
clouds can collide within the jet with Lorentz factors of a few and the 
cloud densities can probably vary by about an order of magnitude. The 
resulting internal shocks move with Lorentz factors of a few through 
the jet. They are thought to be responsible for the observed prompt phase 
of GRBs \citep{Piran:1999jt,Fox:2006vs}. The prompt emissions due to the 
jet thermalization precede the longlasting afterglow, which has its origin 
in the interaction of the jet plasma with the ambient medium.

The underlying mechanisms causing the observed electromagnetic radiation 
are still not fully understood. Highly polarized gamma-ray emission 
suggests the presence of an ordered magnetic field in the emitting zone 
\citep{Steele:2009uq}. These primordial magnetic fields, 
which can be amplified further by the internal shocks, together with 
ultrarelativistic electrons give rise to electromagnetic emissions. The 
resulting photon seed can be upscattered to higher energies by secondary 
processes \citep{Kirk:2010ys}. It is not yet clear if, and to what 
extent, stable and large-scale magnetic fields are generated or amplified 
by instabilities close to internal shocks 
\citep{Medvedev:1999sf,Brainerd:2000rz,Waxman:2006qy}. The presence of 
ultrarelativistic electrons in the jet also cannot be taken for granted. 
The dominant blackbody radiation component of GRB jets suggests a plasma 
temperature of $\sim$100 keV \citep{Ryde:2005rm}. The majority of jet 
electrons are thus only moderately relativistic. The energetic electrons 
responsible for the nonthermal radiation component of the prompt emissions 
\citep{Ryde:2005rm} must consequently be accelerated within the jet, probably 
by the internal shocks. 

The likely involvement of plasma collisions for both magnetic field 
generation/amplification and for electron acceleration and the possible presence of a guiding magnetic field, motivates
our simulation study of the early stages of plasma cloud collision and shock formation.

\subsection{General Plasma behaviour}

The single-fluid magnetohydrodynamic (MHD) approximation can be used
to model the ejection of relativistic jets by compact objects and to
examine their time evolution \citep{Nishikawa:2005yg}. However, it can
not adequately describe the plasma dynamics within the shock transition layer, in
which the ultrarelativistic electrons and the strong magnetic fields
are generated. Many different wave-modes (e.g. upper hybrid waves,
whistler waves and Bernstein mode waves) and instabilities \citep{Bret:2009ec} 
not captured by a MHD model are present and they interplay. The wave and 
instability spectrum depends critically on multiple parameters of the bulk 
plasma, among others the electric and magnetic field orientation and 
strength, the plasma composition, the pre- and post shock density ratio 
and the temperature. In order to model the acceleration of particles, 
the generation of magnetic fields by internal shocks in GRB jets and their 
small scale structure, it is necessary to use a kinetic particle-in-cell 
(PIC) simulation code.

\subsection{Modelling and PIC simulations}

The theory of collisionless magnetised shocks divides naturally into shocks 
with a quasi-parallel magnetic field (treated in this work) and those with 
a quasi-perpendicular shock. Quasi-perpendicular shocks have been well 
studied in the past and in the context of SNR or Solar system shocks 
using both analytical and numerical approaches. Single-fluid analytical 
models have been unable to describe successfully the TeV emission from SNRs 
but hybrid or PIC simulations may provide better insight \citep{Kirk:2001fk}. 
Nonrelativistic numerical models of perpendicular and parallel 
electrostatic shocks have been devised using hybrid methods, for example by 
\citep{Leroy:1981zr,Quest:1988fr}. 
Perpendicular shocks have been modelled by 
\citet{Lee:2004lr,Scholer:2004qy,Amano:2007fj,Umeda:2009rt,Lembege:2009mz} and
many more authors with PIC simulations, while \citet{Sorasio:2006fv} 
addressed fast unmagnetized shocks.
{ Oblique, strongly magnetised shocks have been studied by 
\citet{Lembege:1989lr,Bessho:1999sf,Dieckmann:2008dp,Sironi:2009kt,Shikii:2010lr,Murphy:2010fk}.
}

Considerable work has been done modelling interpenetrating and colliding 
plasma streams in the context of GRB jets, with their attendant wave modes 
and instabilities. \citet{Frederiksen:2004rm} studied using 3D PIC simulations
two initially unmagnetized plasma clouds colliding with a density ratio of 3,
 both composed of electrons and ions. The Lorentz factor of the 
collision speed was 2-3. Computational constraints demanded a reduced 
ion-to-electron mass ratio of 16. The effects of a guiding magnetic field
have been considered by \citet{Nishikawa:2003eu,Hededal:2005zl,Dieckmann:2006tx} 
with 2D and 3D PIC simulations. 

GRB jets may carry with them a significant fraction of positrons 
\citep{Piran:1999jt}. \citet{Kazimura:1998fp,Jaroschek:2004lq,Spitkovsky:2008rm} 
modelled with 2D and 3D PIC simulations the collision of two unmagnetized clouds, 
each consisting of electrons and positrons. \citet{Hoshino:1992fk} introduced heavier 
ions in 1D simulations. Magnetic field effects on such collisions were taken into 
account by \citet{Spitkovsky:2005hi,Sironi:2009kt}. Interpenetrating leptonic 
plasmas have been investigated by \citet{Silva:2003rw}.

The simulation results typically show the generation of magnetic fields by a 
filamentation of the plasma that implies a separation of the currents, provided 
that the guiding magnetic field is not too strong and that the flow speeds are 
relativistic \citep{Cary:1981sf}. The energy density of the magnetic field reaches 
typically about 10\% of the leptonic flow energy. However, most simulation studies 
could not observe a suprathermal population, as long as the plasma cloud collision 
speeds or the beam speeds are mildly relativistic. Such an acceleration of electrons 
on short spatiotemporal scales is conditional on both the presence of ions in the 
plasma flow and a mechanism that can transfer a significant fraction of the ion flow 
energy to the electrons. 

Initial conditions that result in a substantial magnetic field amplification 
and in the electron acceleration to ultrarelativistic energies have been 
proposed by \citet{Bessho:1999sf}, who studied in one spatial dimension the 
collision of magnetized plasmas at the speed 0.9c and with a magnetic field 
direction tilted by 45 degrees relative to the flow velocity vector. An
ion-to-electron mass ratio of 100 was used. The acceleration of electrons up 
to Lorentz factors of $\sim$130 has been found. 

Here we consider the collision of two plasma clouds with a density ratio of 10,
with a speed of 0.9c, and in the presence of a strong magnetic field. The
initial magnetic field would correspond to a primordial jet magnetic field
\citep{Lyutikov:2003vy,Granot:2003mw}. These initial conditions are similar 
to those used by \citet{Bessho:1999sf}. Our magnetic field direction is, 
however, tilted at 0.1 radians relative to the flow direction and, most 
importantly, the two-dimensional simulation permits a more complex array of 
physical processes at a higher mass ratio. A 1D study using almost the same 
parameters \citep{Dieckmann:2008dp} suggests that the magnetic and electron 
energy density will increase drastically in the forming shock transition 
layer. This expectation is confirmed here, but we will demonstrate that the 
actual plasma dynamics in our 2D simulation differs notably from that in the 
previous 1D studies.

{ We demonstrate that the coherency of the circularly polarized electromagnetic 
wave ahead of the shock is reduced compared to what was assumed for the 1D 
simulations, which can be partially attributed to the failure of the guiding 
magnetic field to suppress the filamentation. The shock planarity, which is 
enforced by a 1D simulation, is destroyed here by the development of a flux 
tube. This flux tube is the focus of the research letter \citet{Murphy:2010lr}, hereafter MDD. Here 
we examine in more detail this flux tube and the plasma conditions in the PIC 
simulation that result in its growth.}
\defcitealias{Murphy:2010lr}{MDD}

In this paper we perform a numerical simulation of two plasma clouds colliding, 
using conditions that are probably appropriate for GRBs. In Section \ref{Method} 
we describe the method used, in Section \ref{Results} we present some results and 
finally in Section \ref{Discussion} we discuss the results obtained.


\section{The Numerical Experiments}
\label{Method}

The particle-in-cell (PIC) simulation method has been described in detail 
elsewhere \citep{Dawson:1983dz}. The plasma is represented by an ensemble of 
computational particles (CPs). These CPs correspond to phase space blocks 
rather than physical particles. Consequently, the charge $q_i$ and the mass 
$m_i$ of a CP of the species $i$ do not have to correspond to the equivalent 
of the physical particles it represents. However, the charge to mass ratio 
$q_i / m_i$ must equal that of the physical particle. Then the ensemble 
properties of all CPs representing the species $i$ are an approximation to 
the ensemble properties of the corresponding plasma species, e.g. an electron 
species or an ion species. 

PIC codes approximate the Klimontovich-Dupree equations \citep{Dupree:1963la},
which correspond to the solution of the Vlasov-Maxwell equations by the method
of characteristics. PIC codes can capture all kinetic wave modes and 
instabilities found in collisionless plasma, if the simulation box size and 
the resolution, as well as the statistical plasma representation, are adequate.

We consider the collision of a dense plasma cloud with a tenuous one. The electrons with mass 
$m_e$ and the ions with mass $m_i = 250 m_e$ of the dense cloud are the species 1 and 2, 
respectively. The electrons of the tenuous cloud are species 3 and species 4 denotes the ions 
of the tenuous cloud. We normalise our variables with the plasma frequency of the species 2, 
with the density $n_2$, the charge $q_2$ and the mass $m_2$. The normalization is useful, in 
that it renders the simulation results independent of the plasma density, which is unknown 
for GRB jets. The skin depth of species 2 is $\lambda_2 = c / \omega_{p2}$. The elementary 
charge is $e$. The quantities in SI units (subscript $p$) can be obtained by the substitutions 
$\mathbf{E}_p = \omega_{p2} c m_i \mathbf{E} / e$, $\mathbf{B}_p = \omega_{p2} m_i \mathbf{B}/ e$, 
$\rho_p = e n_2 \rho$, $\mathbf{J}_p = e c n_2 \mathbf{J}$, $\mathbf{x}_p = \lambda_2 \mathbf{x}$ 
and $t_p = t / \omega_{p2}$. The solved equations are

\begin{eqnarray}
\nabla  \times \mathbf{E} =  -\partial_t \mathbf{B}, \,
\nabla  \times \mathbf{B} =  \partial_t \mathbf{E} + \mathbf{J}, \\
\nabla  \cdot \mathbf{B} =  0, \nabla  \cdot \mathbf{E} =  \rho, \\
d_t \mathbf{p}_j = q_j \left( \mathbf{E} + \mathbf{v}_j \times \mathbf{B} \right), \,
\mathbf{p}_j=m_j \Gamma \mathbf{v}_j, \, d_t \mathbf{x}_j =\mathbf{v}_j.
\end{eqnarray}

Here the subscript $j$ refers to the $j^{th}$ CP with the mass $m_j$ and the charge $q_j$. We 
use the Plasma Simulation Code PSC for our simulations, which is a relativistic 3d MPI parallel 
domain decomposed PIC code. It has been extensively used in the laser plasma community 
\citep{Roth:2001wc, Cowan:2004ts}.

\subsection{Simulation Initial Conditions}

    \begin{figure}
   \centering
   \includegraphics[width=\columnwidth]{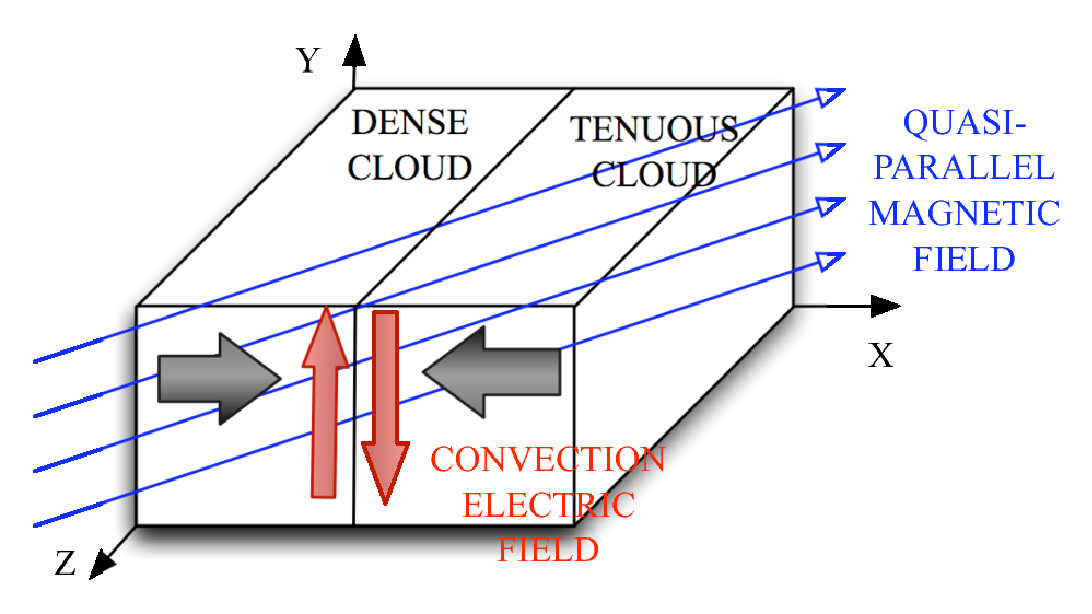}
      \caption{Sketch of simulation initial conditions
                    }
         \label{FigInitScheme}
   \end{figure}
   
We begin the simulation with two colliding plasma clouds. Figure \ref{FigInitScheme} illustrates 
the flow and field geometry. The inclination angle relative to the flow velocity vector of the 
initial magnetic field $\mathbf{B}_0$ is 0.1 radians and its magnitude $|\mathbf{B}_0| = R^{1/2}$ 
in our normalization, where $R =m_i/m_e=250$ is the reduced ion/electron mass ratio. The magnitude
$|\mathbf{B}_0|$ is thus such, that the electron cyclotron frequency equals the electron plasma
frequency of the dense plasma cloud. { Both beams (high and low density) travel initially at  $0.63c$ but in opposite directions, giving a relative speed of $0.9c$.
 } This speed jump will be distributed over the forward and reverse shocks. 
The temperature of all species is $T =131$keV. The thermal velocity of the electrons is $v_{th,e}=
\sqrt{kT/m_e} =0.83 v_b = 0.52c$. The thermal velocity is such that it is not greatly smaller than 
the collision speed, as expected to be for internal shocks in GRB jets. The thermal velocity of the 
ions is $v_{th,i} = v_{th,e}/ {\sqrt R} = 0.033c$. The initial distribution of the particles is a 
relativistic Maxwellian or Maxwell-J\"uttner distribution. The initial density ratio is chosen to 
be 10. 
{ 
We cannot use here the piston method of \citet{Forslund:1970bh}. In this 
piston method, the plasma is reflected at a conducting wall. The plasma 
symmetry across the wall is exploited and only one cloud has to be modelled. 
This cloud is reflected onto itself by the wall and a shock develops at
the collision boundary. It is a computationally efficient method. However, 
only a limited number of field geometries should be modeled with it. The 
electric fields at a conducting wall must point along the surface normal
and the magnetic field orthogonal to it. We can also model with it 
unmagnetized plasma as in \citet{Forslund:1970bh}. It is furthermore 
not possible to let an oblique magnetic field stream with the plasma into 
the simulation box through an open boundary and let the electromagnetic field 
distribution at the conducting wall develop self-consistently when the plasma
reaches it. The oblique geometry implies a flow-aligned magnetic field 
component, which would have to end at the front of the inflowing plasma 
cloud. A magnetic monopole would be the consequence. While the particular 
choice of the boundary condition may not affect the long term evolution of 
the shock far from the boundary, a representation of both colliding plasma 
clouds will be more realistic. In our case study, we collide two plasma 
clouds with different densities and, in fact, this asymmetry rules out the 
piston method altogether due to an absent mirror symmetry at the wall.} 
This is physically 
motivated by the expectation that plasma clouds of similar but unequal density will collide in the 
GRB jet, although we have to point out that the collision boundary is unlikely to be as abrupt as 
the computationally convenient one, which we implement here. However, we do not expect a strong 
dependence of the key simulation results on the boundary shape. The smooth boundary used by 
\citet{Bessho:1999sf} and the sharp one used by \citet{Dieckmann:2008dp} did not result in qualitatively 
different simulation results. The simulation will furthermore show that the important structures form 
well after the initial time, when the boundary has been smeared out. The transport of the magnetic field 
$\mathbf{B}$ at velocity $\mathbf{u}\parallel \mathbf{x}$ gives a convection electric field $\mathbf{E}_c
= \mathbf{u}\times \mathbf{B}$ \citep{Baumjohann:1996ao}. It changes its direction at the cloud collision 
boundary.

\subsection{Linear instability}

The original motivation to pick the high initial plasma temperature, the strong guiding magnetic field 
$\mathbf{B}_0$ and the asymmetric beam density has been to suppress the filamentation instability and,
thus, to enforce a planar shock \citep{Dieckmann:2008dp}. We can test this assumption by computing the 
approximate linear dispersion relation under the following assumptions. An early time is considered, 
when both clouds overlap in a small interval along $x$. The interval is large enough to ensure that the 
electrons of both clouds have mixed and form a single, spatially uniform and hot distribution. The 
interval is small enough so that firstly the ion distribution is unchanged and secondly the magnetic 
field component orthogonal to the beam velocity vector has not yet been compressed to a significant 
amplitude. Then we can approximate the plasma in the cloud overlap layer by two counter-propagating ion 
beams, which move through a hot electron background along a guiding magnetic field with an amplitude 
that equals $250^{1/2}$. The relative speed between both ion beams is 0.9c and their density ratio is 10. 

\begin{figure}
\includegraphics[width=\columnwidth]{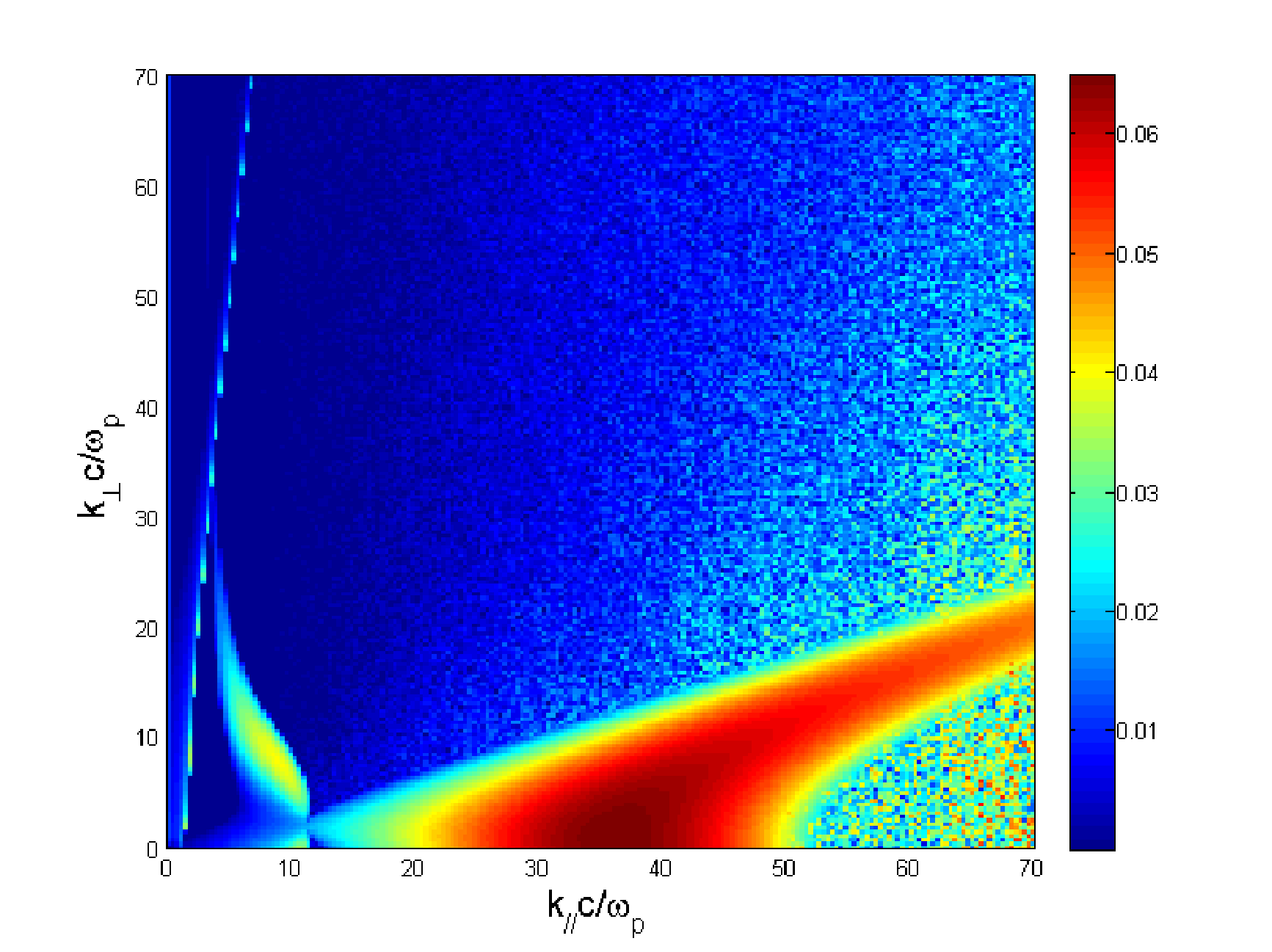}
\caption{The exponential growth rate of the linear instability in units of $\omega_{p2}$ for the ion-to-electron 
mass ratio $R=250$ used in the simulation. The wavenumbers are normalized to the skin depth $\lambda_2$ of the 
dense ion species.}
\label{Dispersion}
\end{figure}

Figure \ref{Dispersion} displays the linear dispersion relation calculated for the ion-to-electron mass ratio of 
$R=250$ in the simulation. 
The exponential growth rates peak in the field-aligned direction, which is characterized by a wavenumber component
along the beam velocity vector $k_\parallel \lambda_2 \neq 0$ and a perpendicular component $k_\perp \lambda_2 = 0$. 
These modes could be observed in a 2D simulation of an oblique shock, which employed a lower collision speed 
\citep{Dieckmann:2010qf}. However, the modes with $k_\perp \lambda_2 \neq 0$ are not suppressed. Bands of unstable 
waves reach out to $k_\perp \lambda_2 \gg 1$. The shock and its downstream region may thus not be planar. 
According to this solution of an idealized linear dispersion relation, the structuring of the shock along its 
boundary is captured well by a simulation box that spans a few ion skin depths into this direction and resolves 
the electron skin depth. Our simulation will represent the wavenumber band $1 \sim k_\perp \lambda_2 < 134$. The 
growth rate map for the mass ratio $R$ is qualitatively similar to that obtained for the correct proton-to-electron 
mass ratio (not shown), which suggests that the spectrum of unstable waves in the PIC simulation will be realistic, 
at least during the initial time. This is not always the case for reduced mass ratios \citep{Bret:2010fj}.

\subsection{Numerical resolution and computational details}
For the 2D simulation, the box measured in ion skin depths is of width $L_x=656 \lambda_2$ and of height $L_y=6
\lambda_2$ resolved in $2.8\times10^{4}$ cells in the propagation direction and 256 cells in the perpendicular 
direction. The total plasma Debye length $L_{T}$, where $\frac{1}{L_T}^2 = \frac{1}{L_{D,ion}^2} + \frac{1}
{L_{D,electron}^2}$ is resolved in 1 cell in the simulation. The electron skin depth is resolved in 2.7 cells. 
{ We use 200 computational particles per cell (100 ions, 100 electrons) in the dense plasma and 100 particle per cell (50 ions, 50 electrons) in the tenuous plasma, 
which is possible by assigning different numerical weights to the CPs. }
No new particles are introduced at the 
periodic boundaries during the simulation. The two plasma clouds rapidly detach from the boundaries. 
{ The high
thermal speed of the electrons implies, that they leave the plasma clouds 
at their rear ends to leave behind a positive net charge. This charge 
separation induces an electric field, which accelerates the ions. This 
process has been researched in the context of a plasma expansion into a 
vacuum \citep{Mora:2009qy}. While such an expansion clearly is an artifact of our initial 
conditions, it does not visibly influence the plasma dynamics at the 
collision boundary. This will be demonstrated below by the supplementary 
movies, which do not show waves or plasma structures that propagate from 
the simulation box boundaries to the cloud collision boundary in the center 
of the simulation box.}
The 
x-boundaries are intentionally placed sufficiently far from the shock forming region that no signal can reach 
the shock forming region traveling from the boundaries within the simulation time.

The number of processors used was 256 Intel Xeon E5462 2.8GHz on an ICHEC 
SGI ICE machine (Stokes). { Total wall clock time was $156 $ hours, giving a cumulative CPU time of $\sim 40,000$ hours for the 
2D simulation.}

%

\section{The simulation results}
\label{Results}

In this section we elucidate the consequences of two plasma clouds colliding. Although the 
plasma dynamics in the cloud overlap layer is determined by both clouds simultaneously,
we will analyse them separately. This is made possible by tagging the CPs initially belonging
to the species under consideration. Henceforth we shall refer with dense ions/electrons to 
the corresponding species of the dense cloud that moves to the right, while the tenuous 
ions/electrons are those of the diluted plasma cloud that moves to the left. In order to 
gain insight into the growth of structures due to the interaction of the clouds, we present 
results of a numerical simulation at the early time of $T1=62$ and the later time of $T2=182$. 
The static images are supported by animated MPEGs available in the online edition, which show 
the time-evolution of the fields.

\subsection{Early stage at time T1}

\begin{figure}
\centering
\includegraphics[width=\columnwidth]{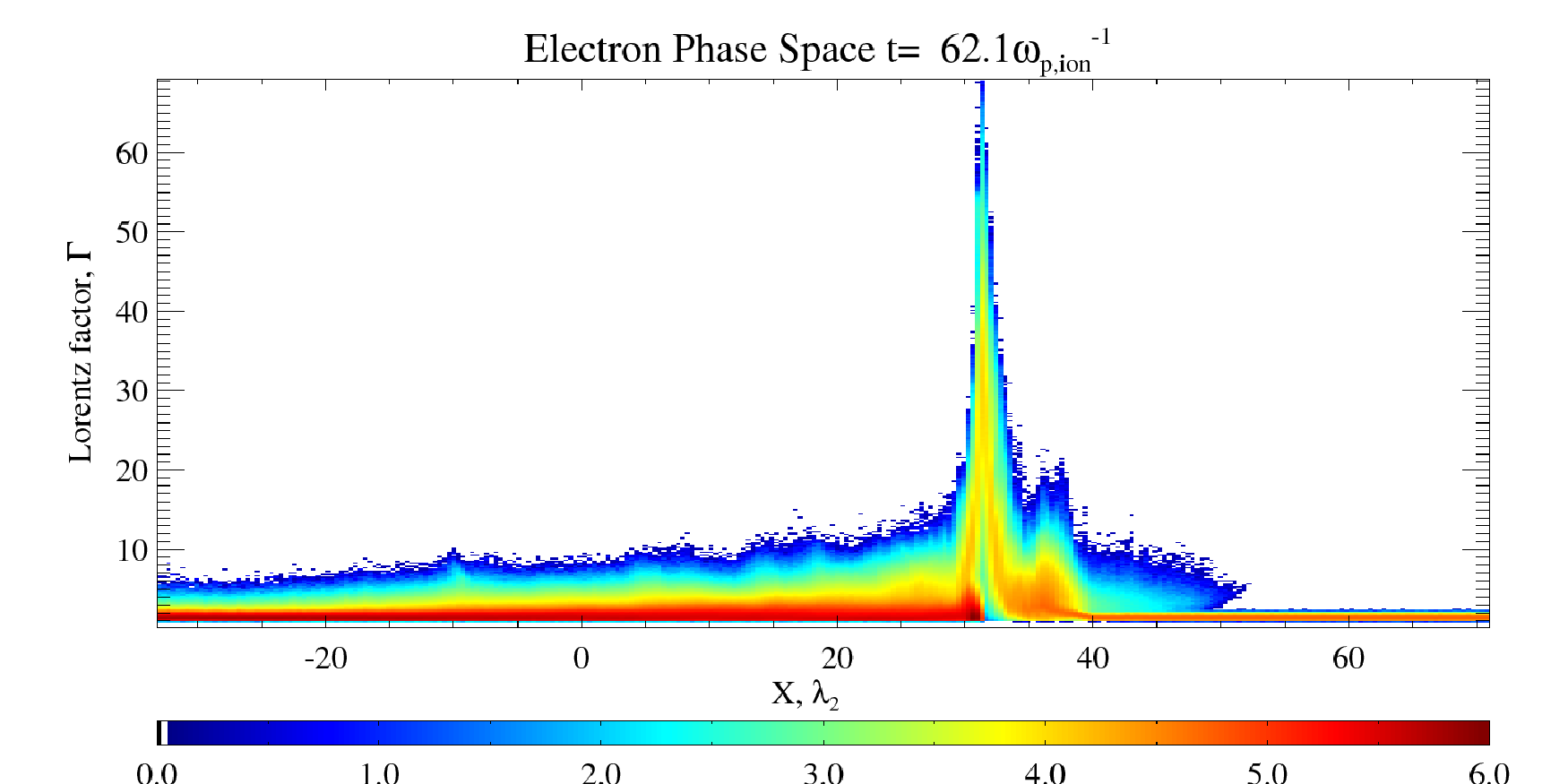}
\caption{Electron phase space: Log of electron density as a function of 
Lorentz factor $\Gamma$ and x at time $t=T_1$.}
\label{FigElecPS1}
\end{figure}

\begin{figure}
\centering
\includegraphics[width=\columnwidth]{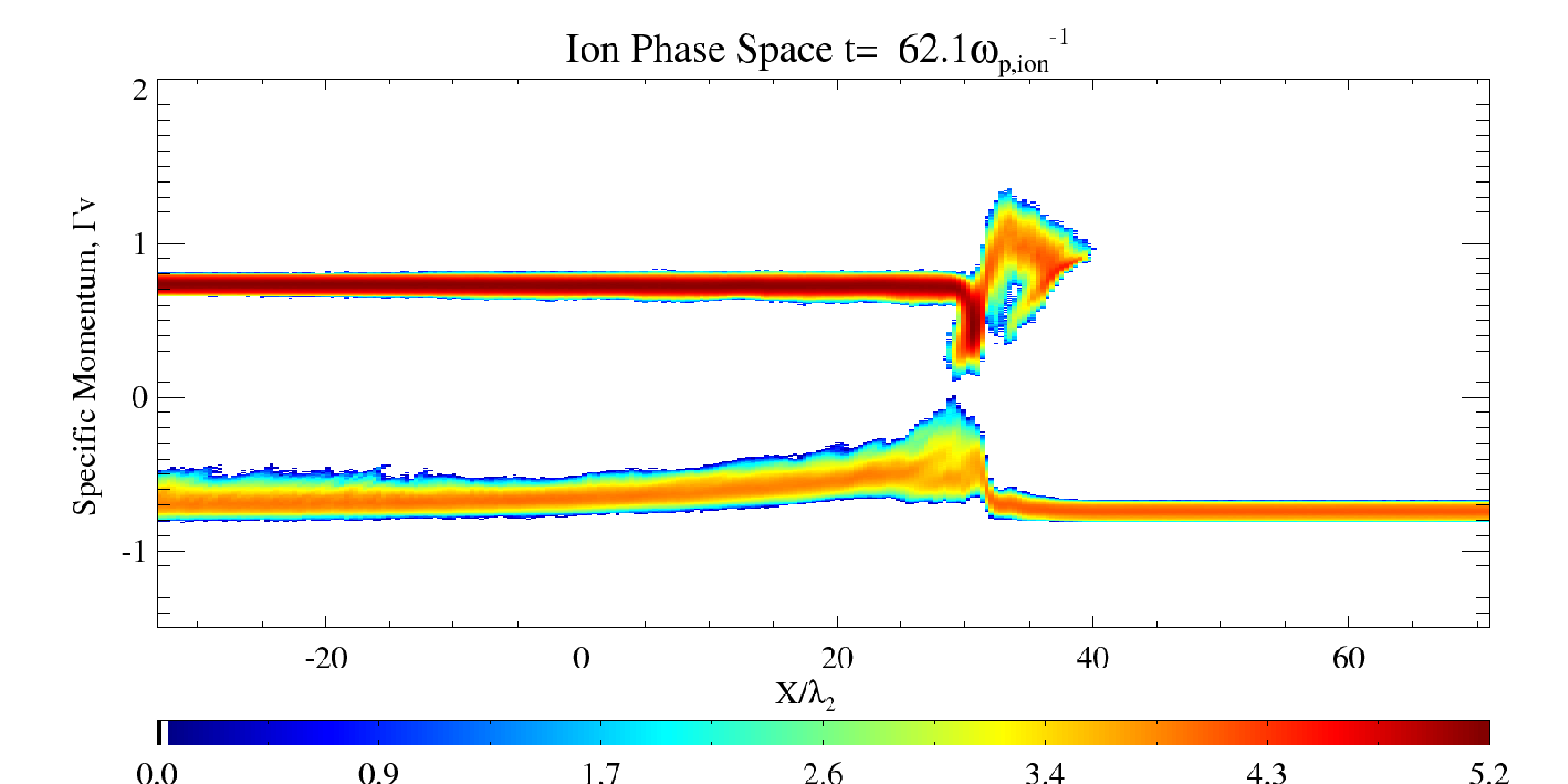}
\caption{Ion phase space: Log of ion density as a function of specific 
x-momentum $\Gamma v_x$ and x at time $t=T_1$} \label{FigIonPS1}
\end{figure}

At the time T1, the beams have counterstreamed for tens of ion skin depths.
We look first at the electron phase space distribution integrated over $y$. 
The dynamics of the electron beams show that the electrons at position x=30 
are already accelerated to $\Gamma=60$ from the initial $\Gamma \le 8$ 
(Figure \ref{FigElecPS1}). The strongest electron acceleration occurs 
for $30<x<33$. The relativistic inertia of these electrons is already 
comparable to that of the ions and we expect a detectable reaction from them.
From the ion phase space plot in Fig. \ref{FigIonPS1}, which is also integrated 
over $y$, it can be seen that little interaction has taken place in between 
the colliding clouds of ions over most of the displayed $x$-interval. However 
modulations of the specific x-momentum of both clouds are highly evident at 
$30<x<40$. All ions are decelerated at $30<x<33$ in the simulation frame, 
which provides a reservoir of energy for the acceleration of electrons and dense ions to $\Gamma v_x \approx 1.2$ at $x\approx 32$. The dense 
ions are decelerated also at $x>35$. 

\begin{figure}
\centering
\includegraphics[width=\columnwidth]{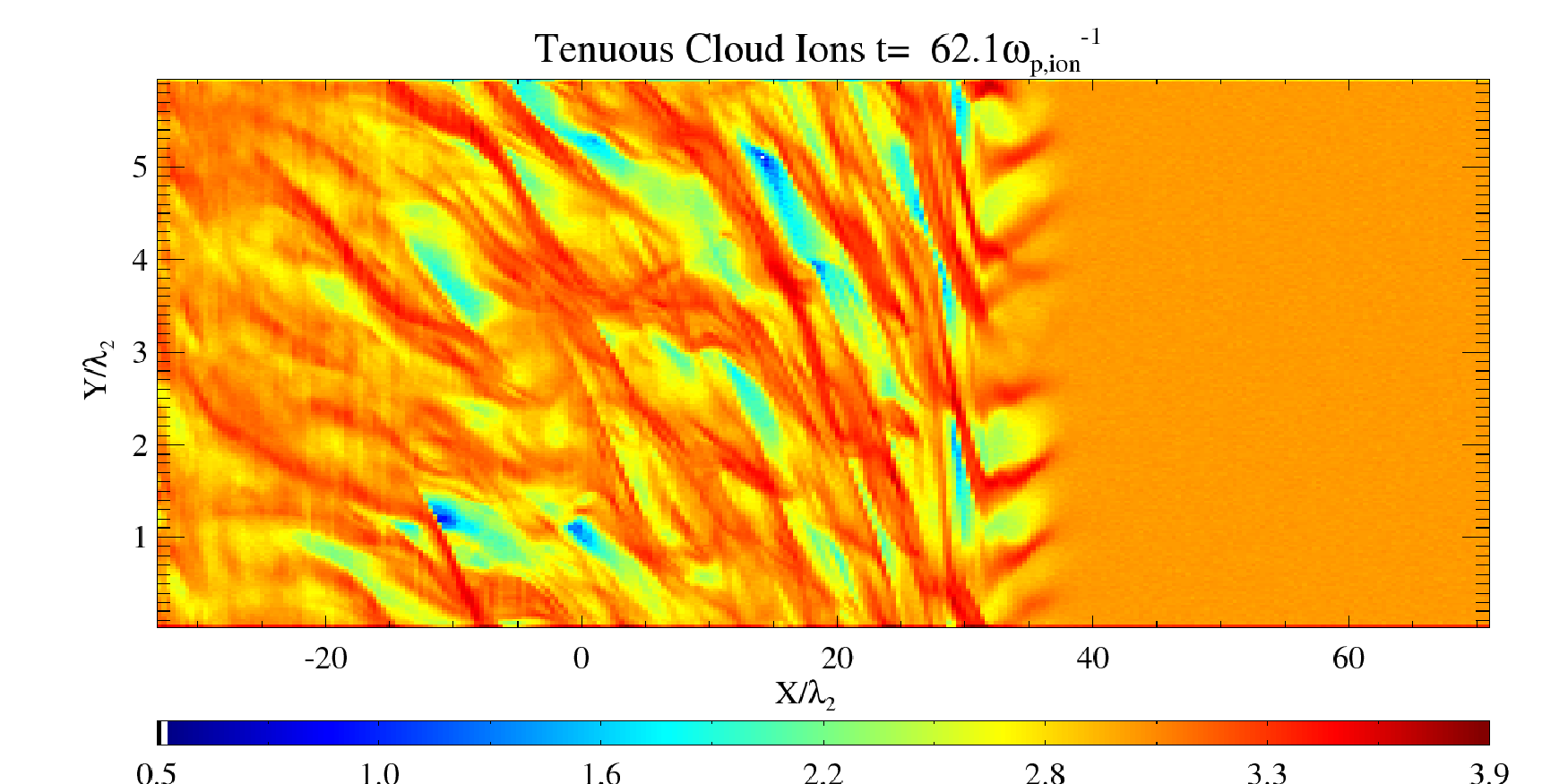}
\includegraphics[width=\columnwidth]{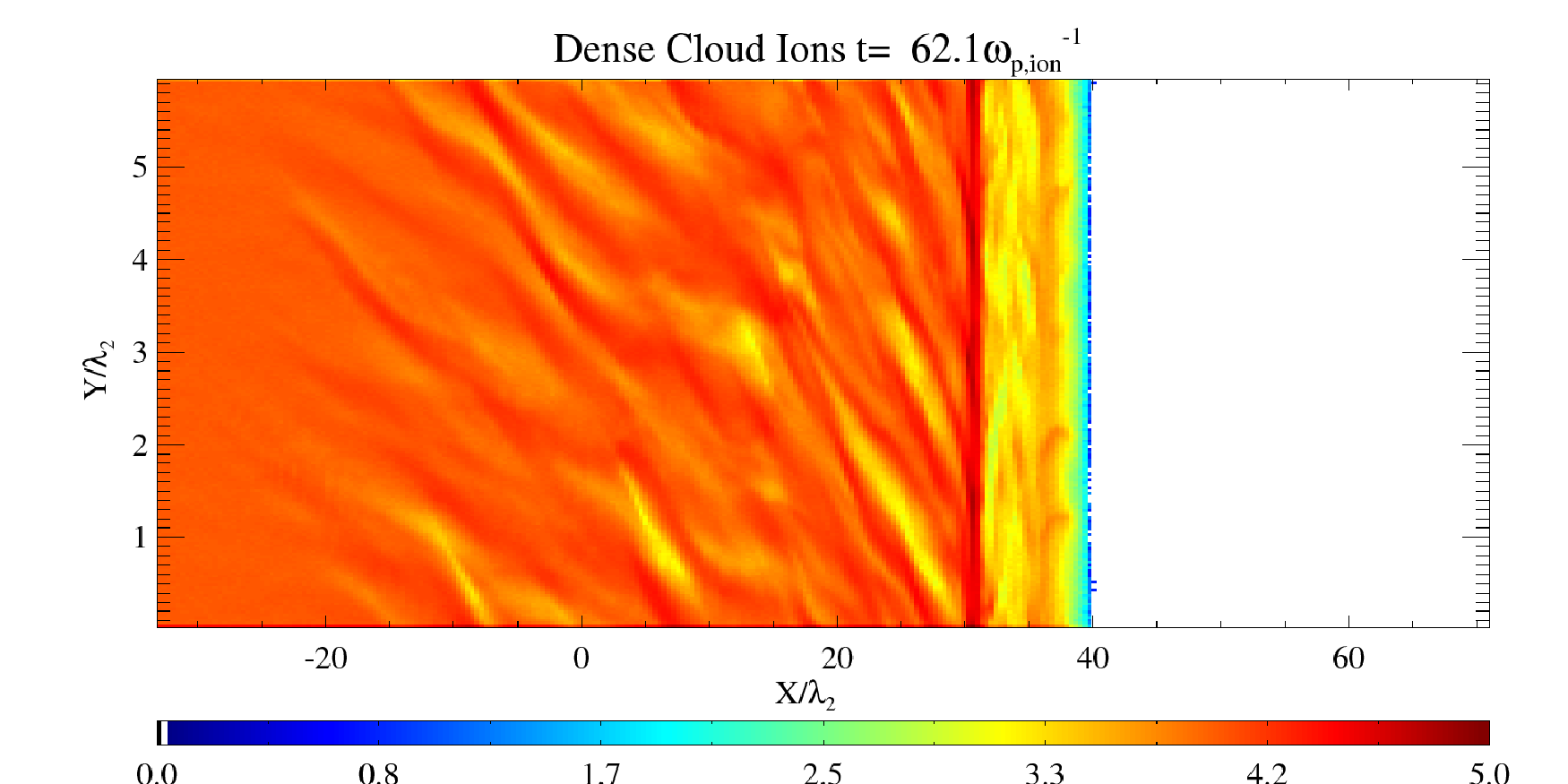}
\caption{Two dimensional logarithm of the ion density as a function of X 
and Y at the time $t=T_1$. The upper panel shows the left moving tenuous
ions and the lower one the right-moving dense ones.}\label{FigIonLeftRightT1}
\end{figure}

The large-scale distribution of the ion densities is shown in Fig. 
\ref{FigIonLeftRightT1} for the tenuous ions and dense ions. The tenuous ions 
undergo a rapid filamentation, as they reach the cloud overlap layer at 
$x\approx 40$. The tenuous ions initially form 
filaments with a thickness of $\approx 0.1$ that are almost aligned 
with the flow velocity vector. 
Assuming the current channels are engendered by
a filamentation instability between the ions of both clouds without electron 
involvement, one should see such structures also in the dense ions. 
This is because both clouds have approximately the same density and temperature for 
$30<x<40$ and both should thus behave similarly. 
Filamentary structures 
resembling those in the tenuous ions are, however, not visible in the dense 
ions on this scale. This filamentation instability must thus involve all 
plasma species. The electrons in this interval still carry a substantial 
directed flow energy, which can be released and further modify the instability. 
At this advanced simulation time, the linear dispersion relation discussed above would no longer be a good 
approximation. 
However, the key conclusion 
we draw from it, namely that the filamentation instability is not suppressed, 
is supported by the ion distribution.

The current channels in the tenuous ions are then rapidly deflected and 
thermalized, as they enter the interval $30<x<33$ in Fig. 
\ref{FigIonLeftRightT1}, 
where the electron acceleration and, thus, the electromagnetic fields are 
strongest. A structured and beamed distribution is present for $x<30$. The 
ions in these channels have not decreased significantly their $p_x$ momentum 
(Fig. \ref{FigIonPS1}). A two-dimensional density modulation is also visible 
in the dense ions for $-20<x<30$. Its density peaks at $x=30$, which is the 
interval where the dense ions are slowed down most in Fig. \ref{FigIonPS1}. 
The planar front of the dense cloud is surrounded by filamentary 
distributions, as it has also been observed for a reduced collision speed 
\citep{Dieckmann:2010qf}.

\begin{figure}
\centering
\includegraphics[width=\columnwidth]{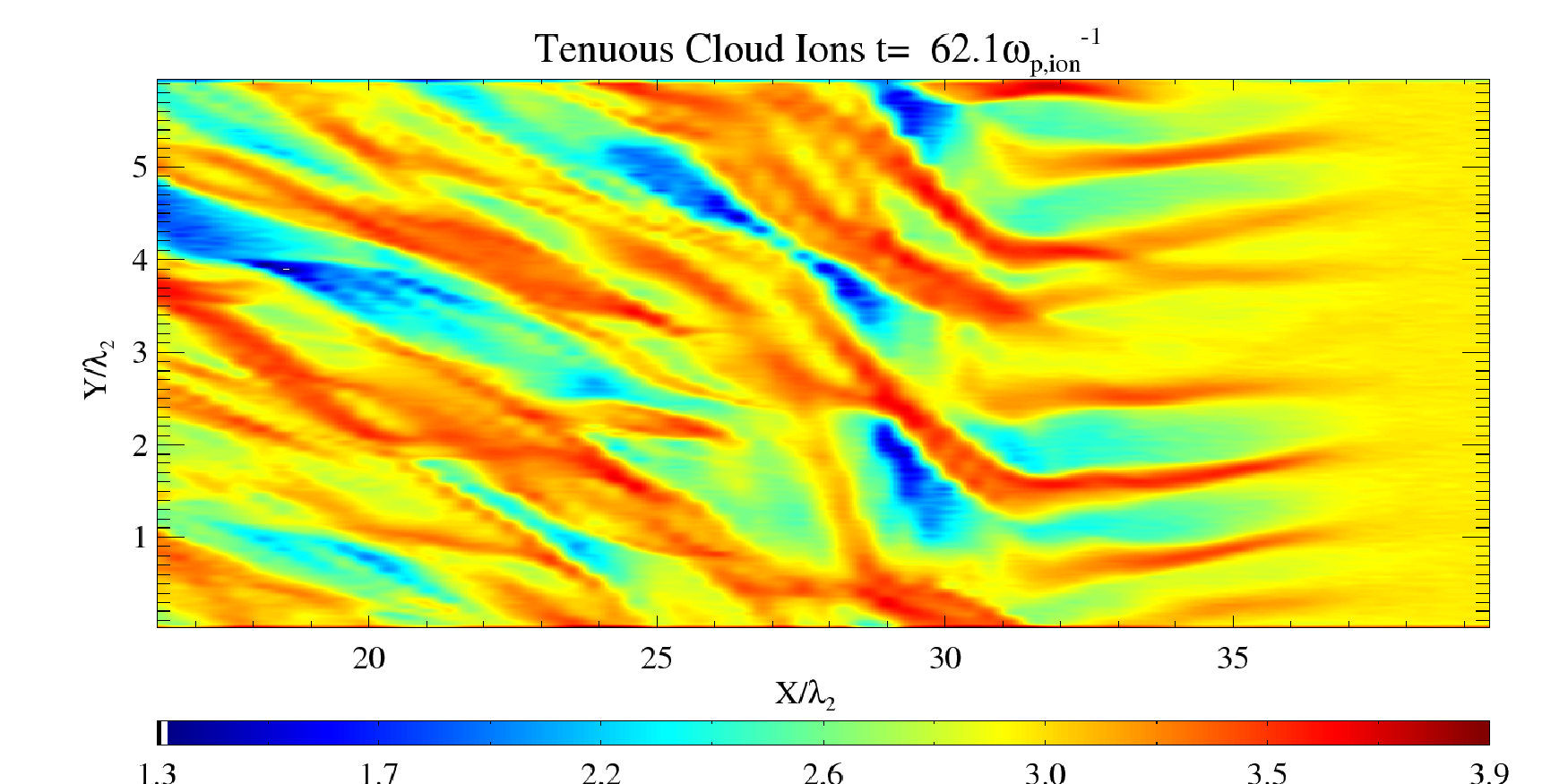}
\includegraphics[width=\columnwidth]{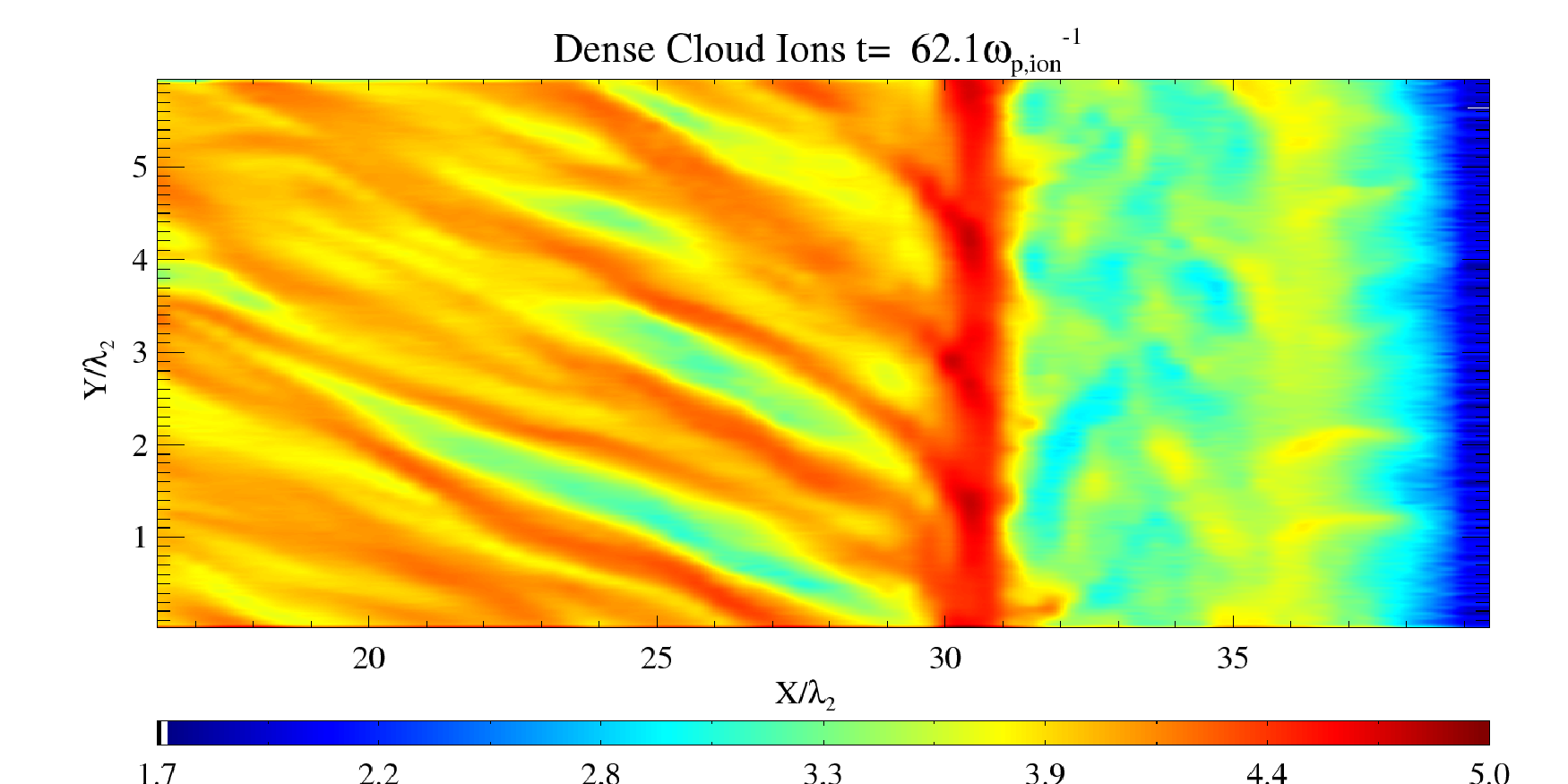}
\includegraphics[width=\columnwidth]{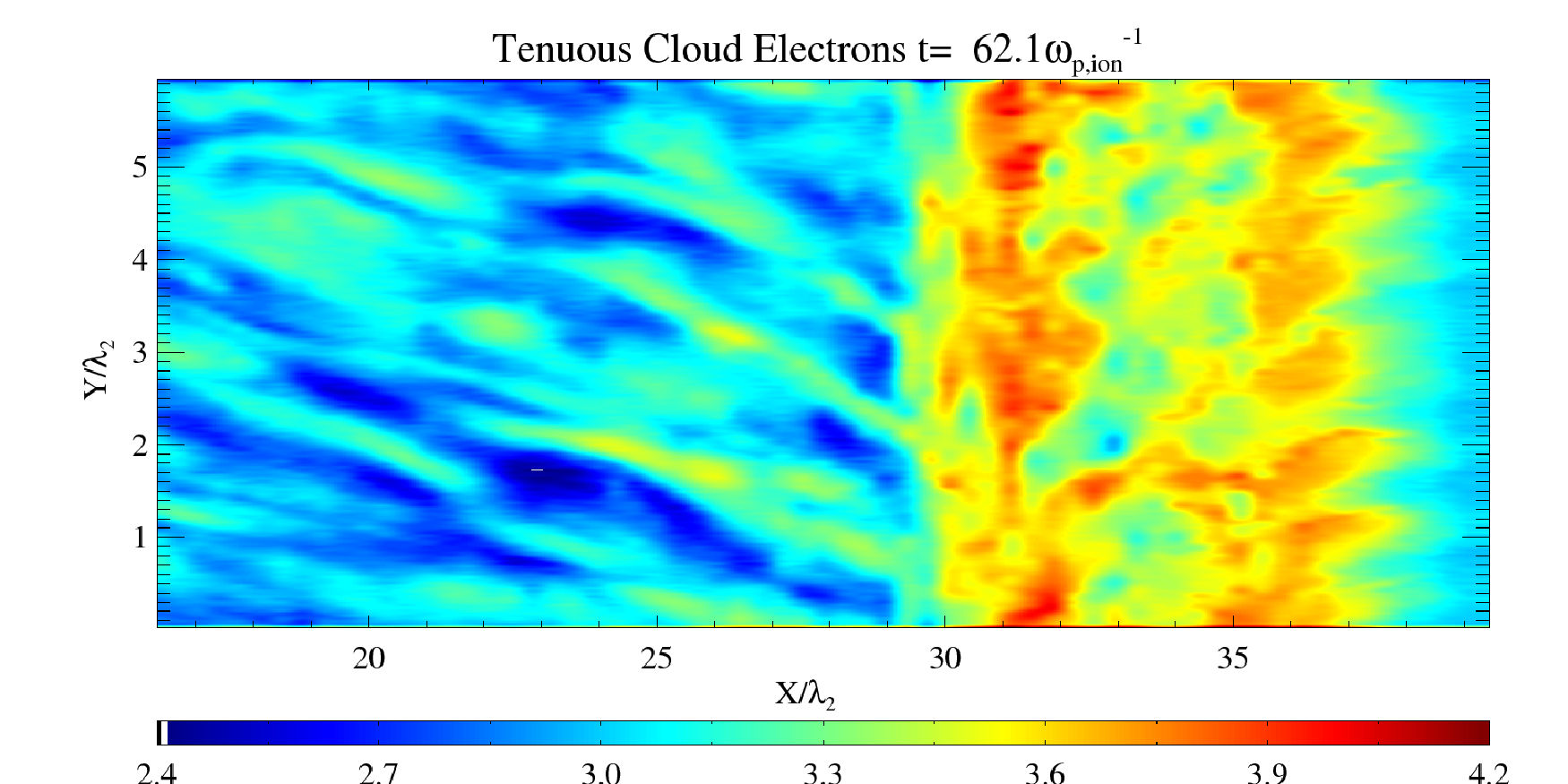}
\includegraphics[width=\columnwidth]{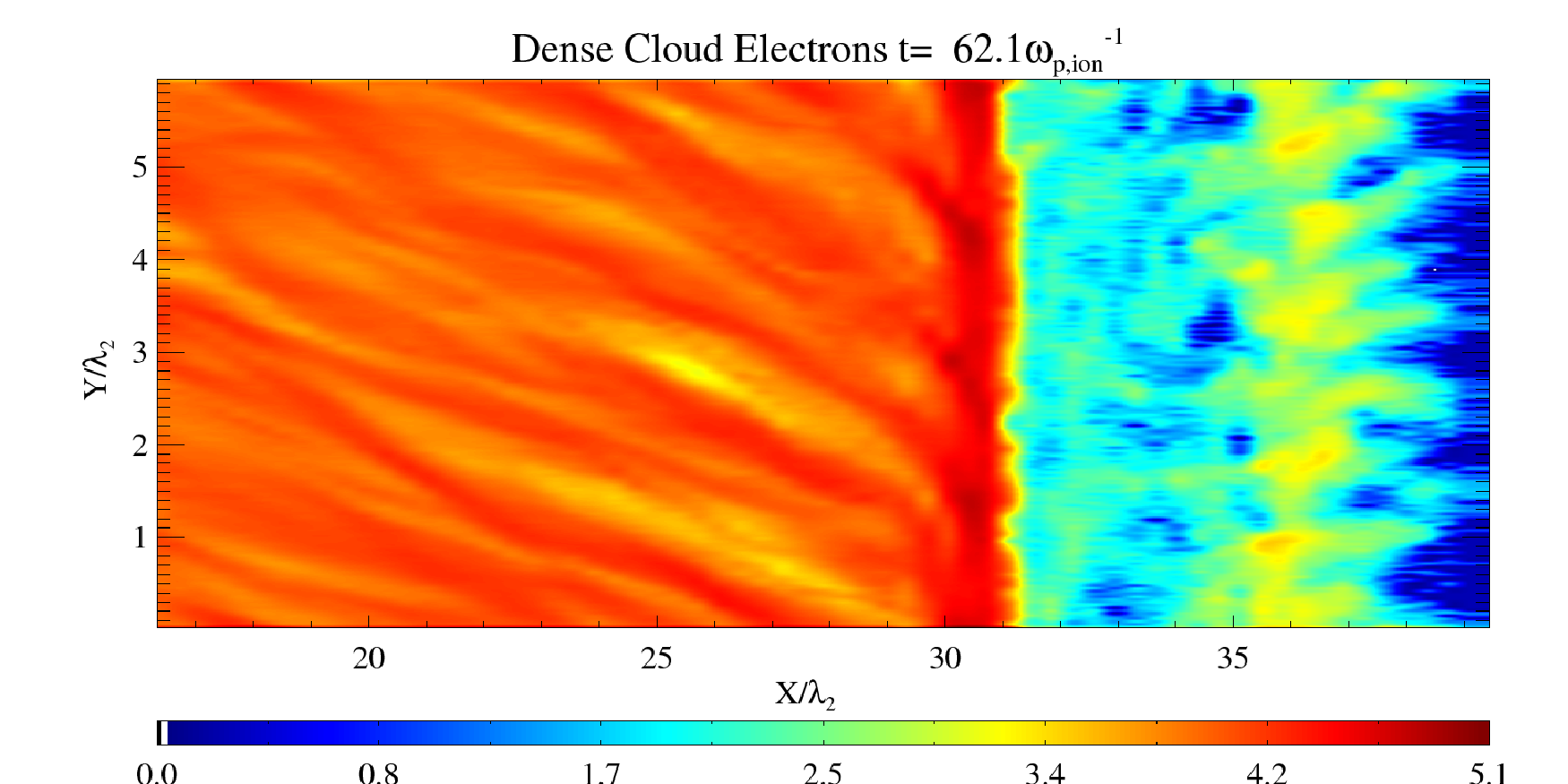}
\caption{Two dimensional plots of logarithm of dense and tenuous ion and electron clouds at time T1, 
for the zone close to the collision boundary.}
\label{FigZoomDen}
\end{figure}

A zoom of the zone around the collision boundary (Figure \ref{FigZoomDen}) 
shows the spatial density distribution of each of the four species. The 
current channels in the tenuous ions start to form at $x\approx 38$. The 
current density increases as we go from this position to $x\approx 34$ and 
the space between the channels is progressively depleted of ions. The current
channels merge, e.g. at $x\approx 33$ and $y \approx 4$. Only 4 major 
current channels eventually cross the electron acceleration region and reach 
$x\approx 30$, where they are deflected to increasing values of $y$ and 
scattered at $x\approx 27$. Remarkable density modulations of the tenuous 
ions are visible at $x\approx 30$, where we find a minimum density of 
$\exp{(1.3)} \approx 4$ at $y\approx 1.7$ and a maximum density of
$\exp{(3.9)} \approx 50$ at $y\approx 2.5$. Their density minima at $(x,y) 
= (29.5,1.7)$, $(30,3)$ and $(30,5.5)$ are correlated with local maxima of 
the density of the dense ions. Some correlations between the filaments of 
the tenuous and the dense ions are visible, e.g. at $x\approx 37$ and 
$y\approx 1$ and $y\approx 2$, where the current channels both have a 
density of $\approx \exp{(3.5)}$. 

The dense ions show a quasi-planar front at $30<x<31$ with an enhanced 
density as well as a less pronounced second front at $x\approx 36$. In 
particular the front at $x\approx 31$ is sharp and the densities of the 
dense ions and electrons decrease by a factor $\approx 5$ and $\approx 20$, 
respectively, as we cross its boundary to the upstream (larger $x$). The 
lower density of the dense electrons ahead of the front is probably caused 
by their rapid expansion upstream. Figure \ref{FigElecPS1} reveals that hot 
electrons leak out of the acceleration region and reach a $x\approx 50$. 
Their current induces a return current in the tenuous electrons, which 
accelerates the latter towards the density peak at $x=31$ and results in 
their accumulation at $30<x<33$ and at $35<x<38$. The dense electrons 
also accumulate in $35<x<38$. The density maxima of the tenuous electrons 
correlate well with the locations, where the electron acceleration in Fig. 
\ref{FigElecPS1} and the ion deceleration in Fig. \ref{FigIonPS1} is 
strongest. The ion deceleration in the interval $35<x<38$ causes their 
accumulation and also that of the electrons, which must cancel the ion 
charge. Furthermore, a strong filamentation along $y$ is visible for 
$20<x<30$ for all four species in Fig. \ref{FigZoomDen}.

\begin{figure}[t]
   \centering
  \includegraphics[width=\columnwidth]{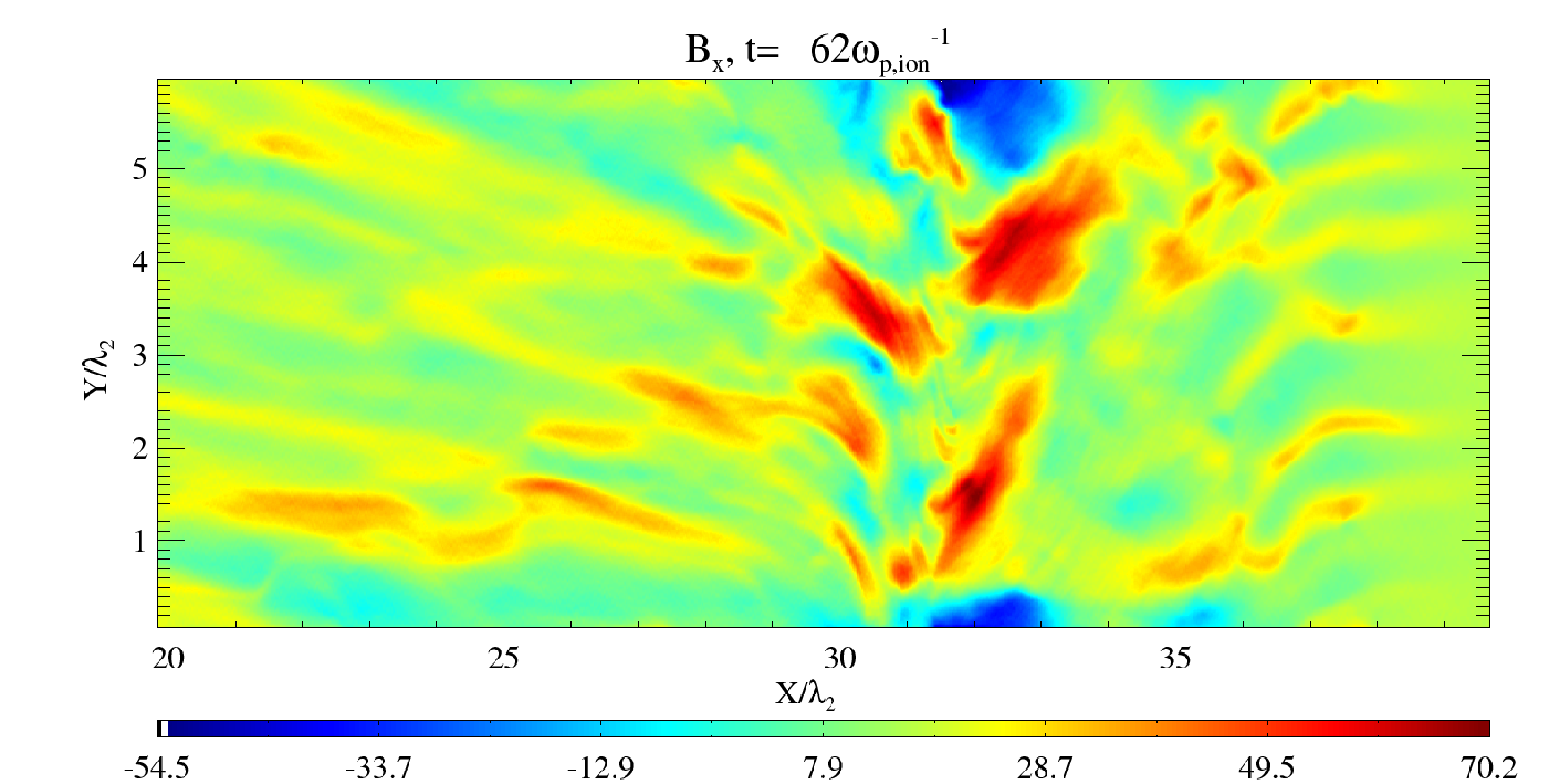}
  \includegraphics[width=\columnwidth]{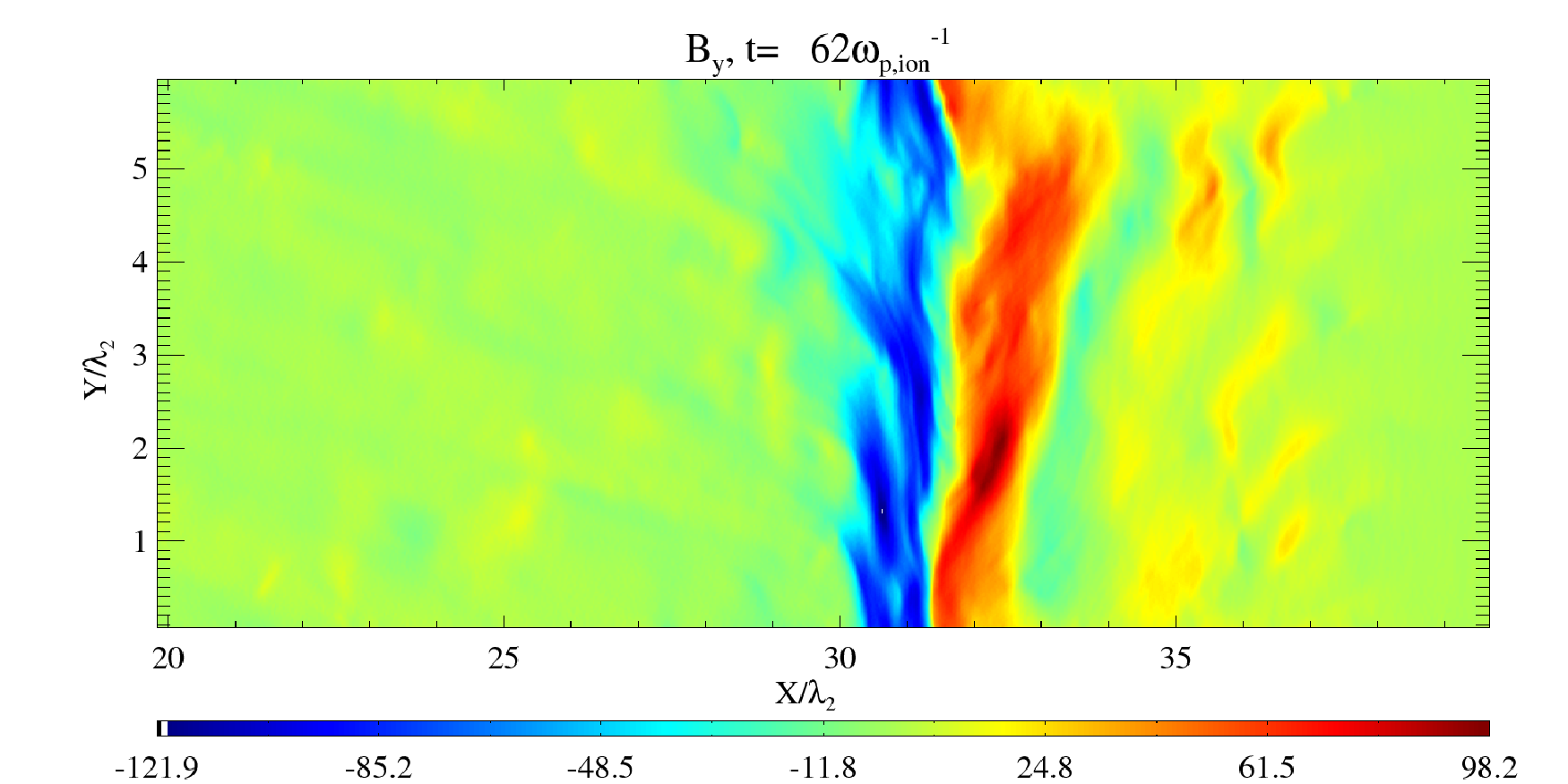}
  \includegraphics[width=\columnwidth]{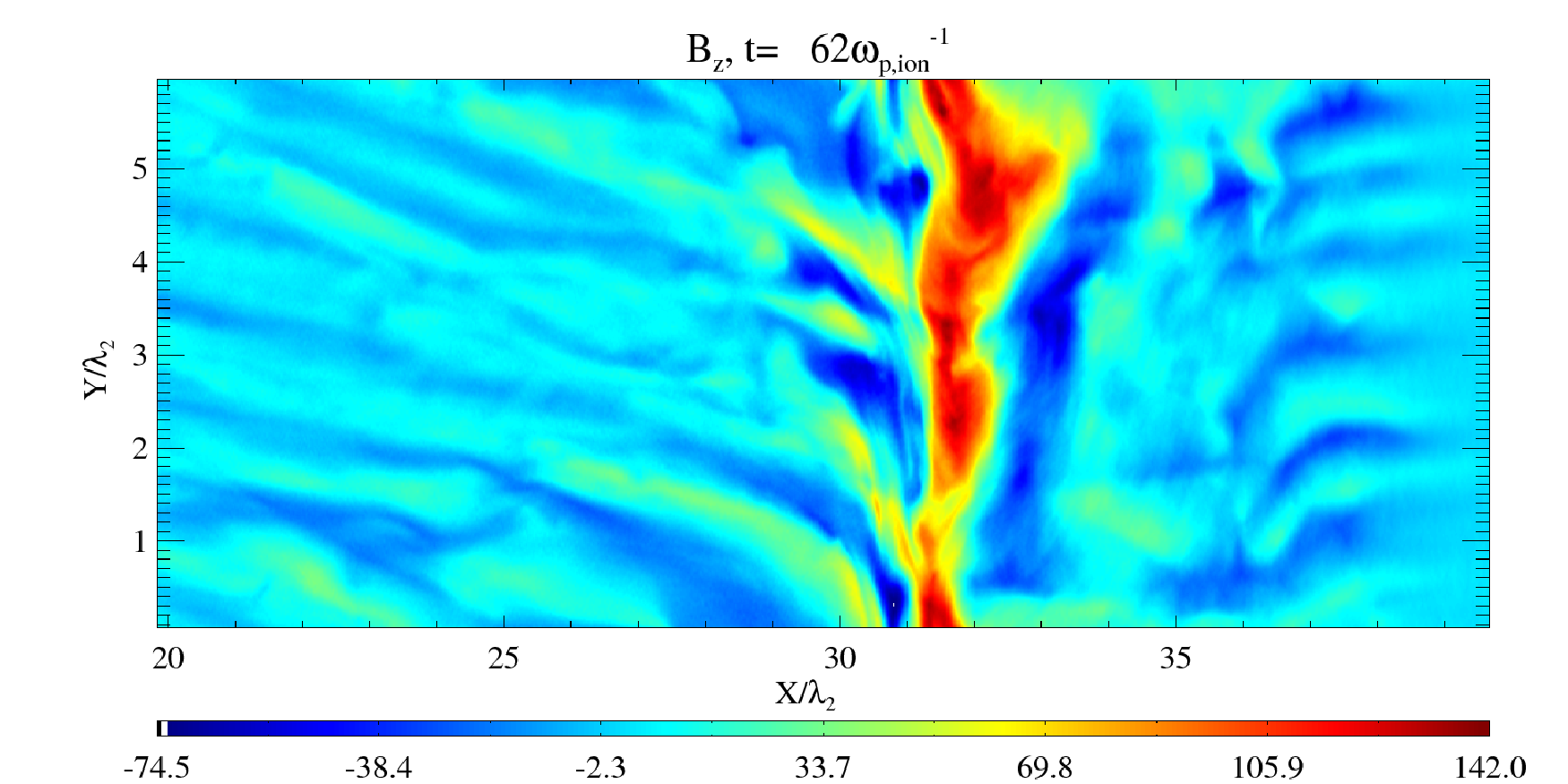}
      \caption{Magnetic field components:  Zoom of 2D linear colourscale plot 
of spatial distribution of $B_x$ (upper plot), $B_y$ (middle plot) and $B_z$ 
(lower plot) at $t=T_1$.}
         \label{FigZoomB}
   \end{figure}

\begin{figure}[t]
   \centering
  \includegraphics[width=\columnwidth]{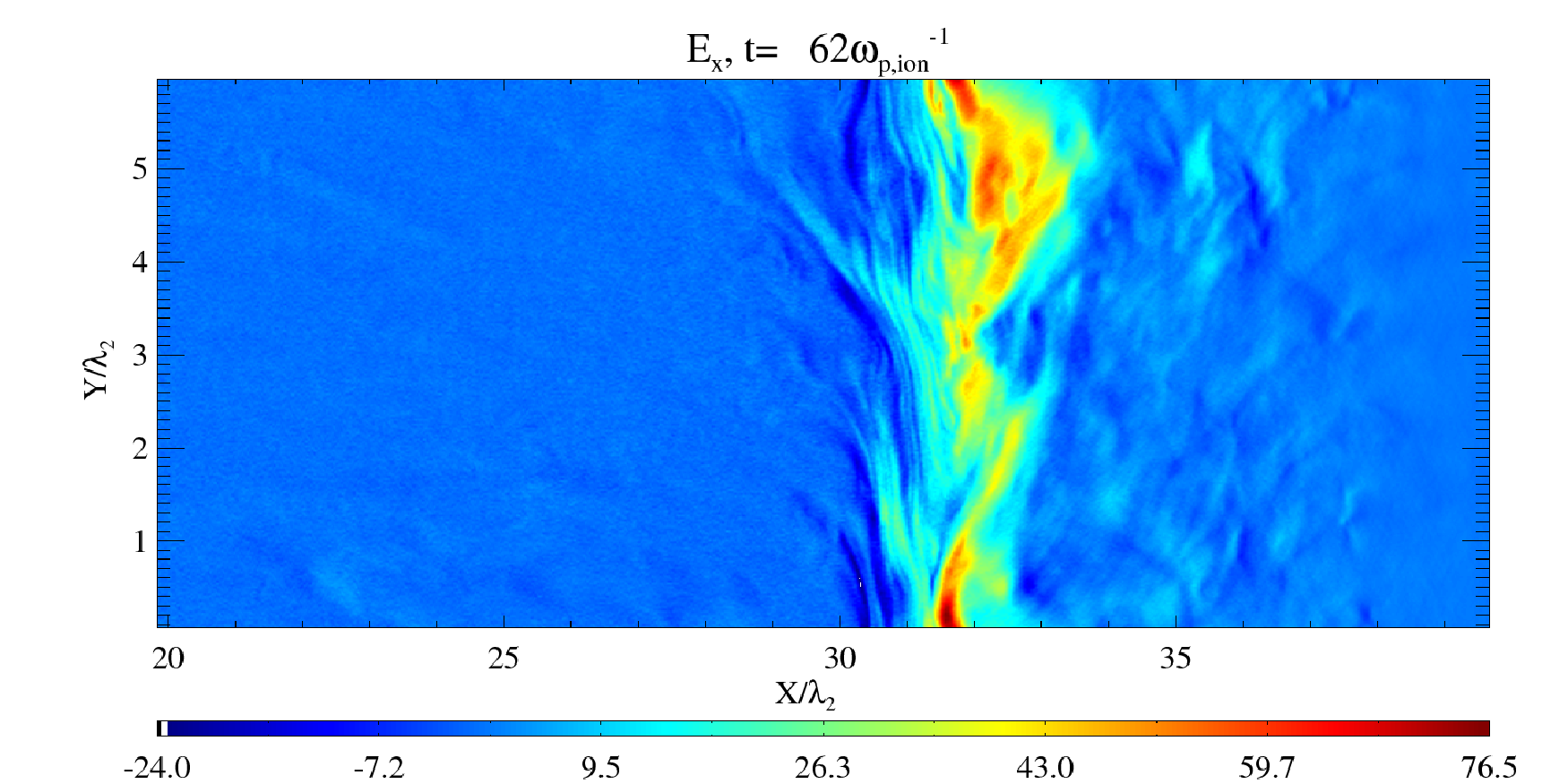}
  \includegraphics[width=\columnwidth]{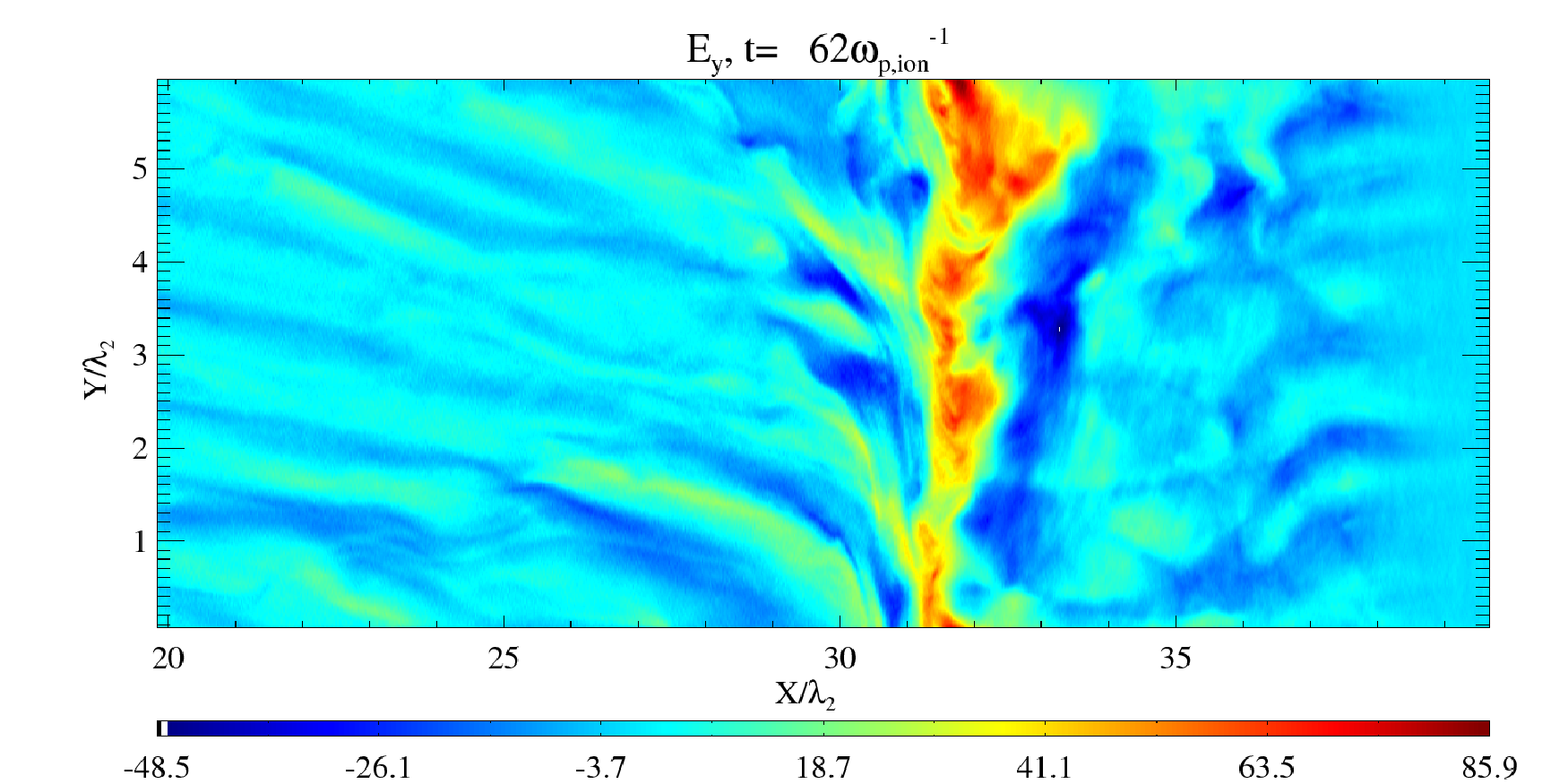}
  \includegraphics[width=\columnwidth]{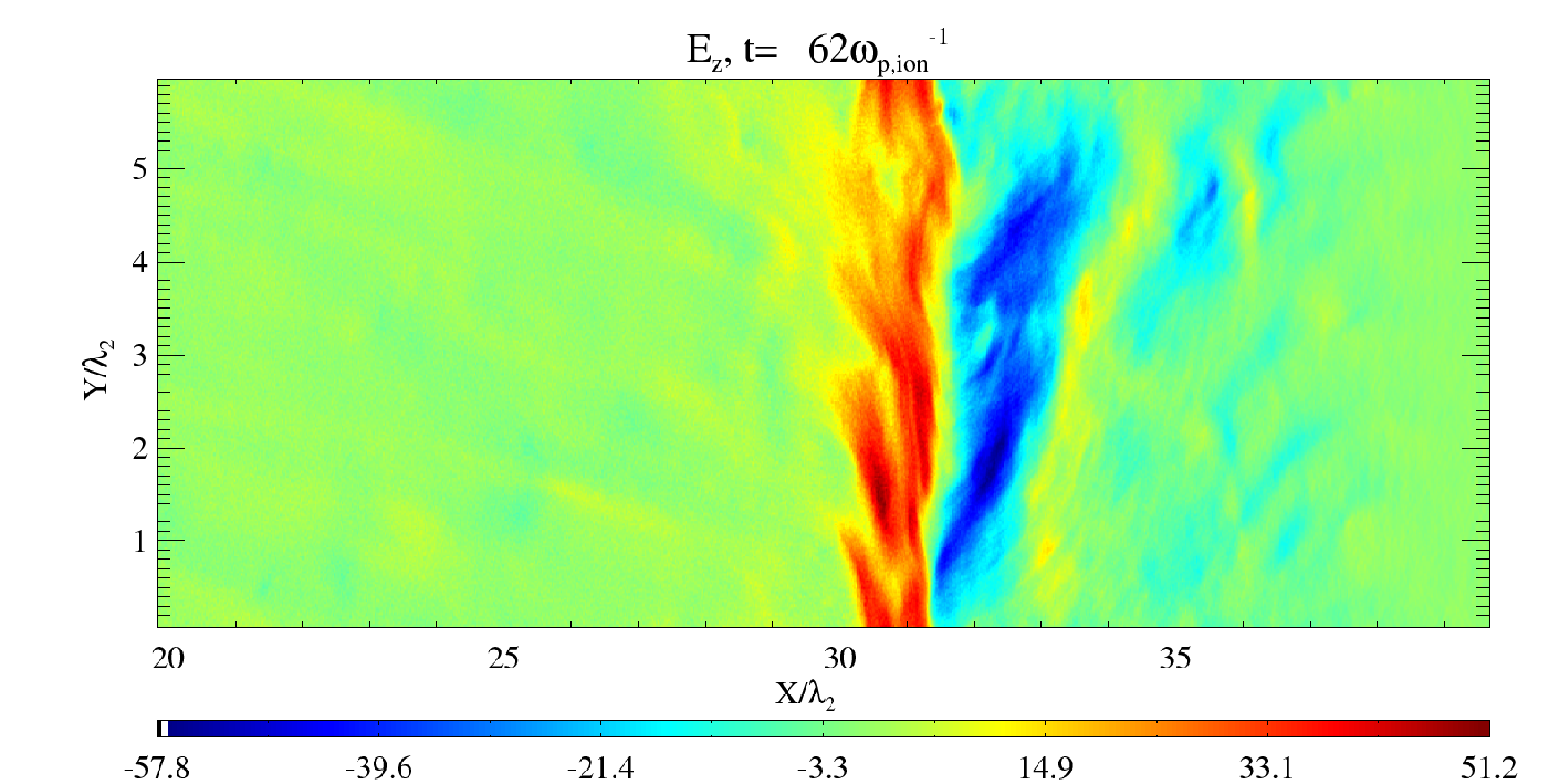}
      \caption{Electric field components: Zoom of 2D linear colourscale plot 
of spatial distribution of $E_x$ (upper plot), $E_y$ (middle plot) and $E_z$ 
(lower plot) at $t=T_1$.}
         \label{FigZoomE}
   \end{figure}

The mechanisms that accelerate the electrons at the expense of the ion 
energy can be identified with the help of the electromagnetic fields. The 
magnetic and electric components in the location of the collision boundary 
(Figures. \ref{FigZoomB} and \ref{FigZoomE}) reveal a large electromagnetic 
pulse in the interval $30<x<33$, which is the interval with the strongest 
electron acceleration and ion deceleration in Figs. \ref{FigElecPS1} 
and \ref{FigIonPS1}. The electromagnetic fields orthogonal to the collision 
(x) direction reveal bipolar pulses, which are shifted in space. This is 
most evident for $B_y$, where the negative and positive magnetic field is 
separated at $x\approx 31.5$. This zero crossing of $B_y$ is approximately 
where $B_z$ reaches its maximum positive value, evidencing a phase shift
of $90^\circ$ along $x$ between $B_y$ and $B_z$. A negative $B_z$ is then 
visible at $x\approx 33$, just ahead of the positive interval of $B_y$. 
The $B_y$ and $B_z$ reach moduli $\approx 100$ each, which is about 6 times 
larger than $|\mathbf{B}_0| = 250^{1/2}$. 

The bipolar pulse of $E_y$ is in 
phase with that of $B_z$, while the pulse in $E_z$ is in antiphase with 
that in $B_y$. The electric field pulse is to some degree the consequence 
of the rapidly moving magnetic field pulse (convection electric field) and 
this contribution would vanish in the rest frame of the magnetic field 
structure moving with the speed $\approx v_b$. The $B_y$ and $E_z$ show a 
further oscillation at $x\approx 35$, suggesting that it is the same 
circularly polarized energetic electromagnetic structure that was observed 
previously by \citet{Dieckmann:2008dp,Dieckmann:2010qf}, although here its 
coherency along $y$ is low and it does not extend far upstream. The low 
coherency can at least in parts be attributed to the strong spatially 
varying $B_x$. The presence of a strong nonplanar $B_x$ implies spatially
varying strong currents in the z-direction.

A single circularly polarized purely electromagnetic wave cannot by itself 
accelerate electrons to ultrarelativistic speeds. Its combination with 
the electric $E_x \gg 0$ in Fig. \ref{FigZoomE} is necessary for this 
purpose. In our normalization the ratio of the values for $\mathbf{E}$ and 
$\mathbf{B}$ is almost equal to the force ratio for a particle, which moves 
with $v\approx c$. The electric and magnetic forces on relativistic particles 
are thus comparable. The $E_x>0$ can be explained as follows. As the upstream 
plasma impacts on the strong magnetic field at $x\approx 33$, the electrons 
are deflected away from their original flow direction, while the ion reaction 
is much weaker. A current develops in the y-z plane, which amplifies locally 
the magnetic field. The electrons fall behind the ions, because their velocity 
along $x$ is reduced. An $E_x>0$ builds up, which tries to restore the 
quasi-neutrality. The tenuous electrons are dragged by it across the magnetic 
field. This cross-field transport accelerates the electrons to relativistic 
speeds, as it is confirmed in Fig. \ref{FigElecPS1}. This acceleration 
acts in the y-z plane, which partially explains the spatial confinement along 
$x$ of the accelerated electrons. This spatial confinement of the strong 
electromagnetic fields and of the electron acceleration also implies a 
localized fast decrease of the ion flow speed, explaining the ion reaction 
within $31<x<35$ in Fig. \ref{FigIonPS1}. The electric field, which drags 
the electrons to the left in this interval, causes the slowdown of the tenuous 
ions as they move to the left and the speedup of the dense ions, which move to 
the right. The consequent ion accumulation further confines with its 
massive positive charge the electrons. This electron acceleration mechanism 
ceases to work, when the $E_x$ is strong enough to stop the ions. 

We can also observe filaments in particular in $B_z$ for $x<30$, which results
out of the filamentary strucures observed in all plasma species in this 
interval. We expect this magnetic component to show the strongest modulation
if the currents are approximately aligned with the $x$-direction and are
modulated along $y$. Magnetic field stripes aligned with $x$ are also observed
for $x>35$. These filaments are driven by the upstream electrons and the 
electrons that leak to higher $x$, reaching $x\approx 50$ in Fig. 
\ref{FigElecPS1}. This upstream filamentation has also been observed at the
faster shock modelled by \citet{Martins:2009ly}.

\subsection{Late stage}
\label{Latestage}

The physics at time
T1 has given us insight into the plasma processes that take place initially
and precondition the plasma such that the vortex forms, which is discussed 
in detail in \citetalias{Murphy:2010lr}. We now consider the simulation time T2, 
which is approximately that investigated in \citetalias{Murphy:2010lr}. Here we 
discuss in more detail the electromagnetic fields at the shock front, which 
have not been shown in \citetalias{Murphy:2010lr}. The fields may not provide more
information related to the vortex than the current, but they are essential
to understand the particle acceleration, which is the focus of this paper.

\begin{figure}
   \centering
   \includegraphics[width=\columnwidth]{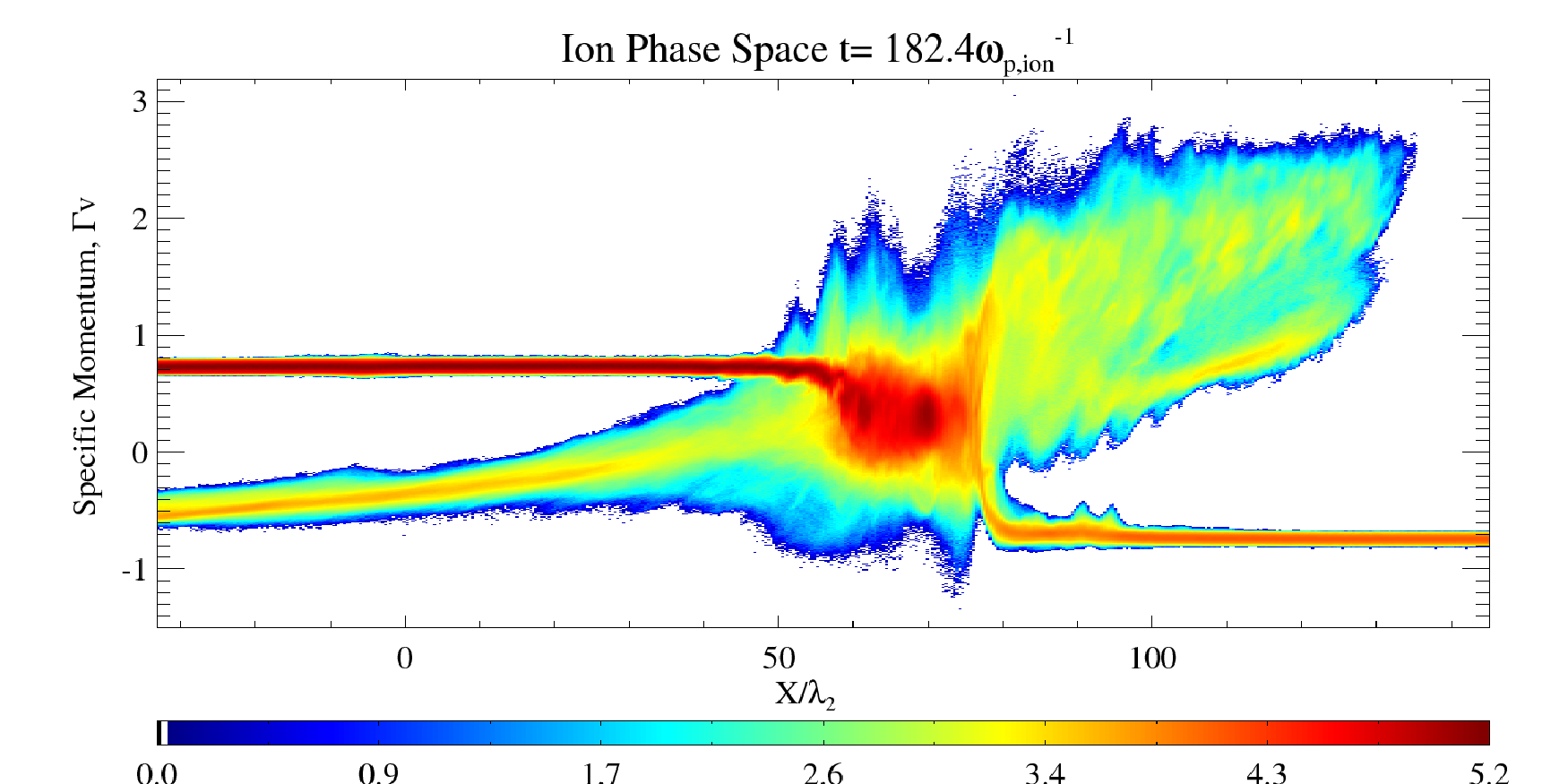}
      \caption{Ion phase space: Logarithm of electron density as a function of specific x-momentum $\Gamma v_x$ 
and x at $t=T_2$. 
                    }
         \label{FigIonPST2}
   \end{figure}

 To aid the analysis, in the simulation three distinct 
zones are identifiable at T2. These zones are most easily distinguished in Fig. \ref{FigIonPST2}, which 
displays (as in Fig. \ref{FigIonPS1}) the ion phase space density integrated over $y$ as a function 
of $x$ and $p_x$, but now at the time $T_2$. The displayed x-interval can be subdivided into the interpenetrating
ion beam zone (IIBZ) with $-35 < x < 60$, the downstream region with $60 < x < 80$ and the foreshock region of
the strong forward shock with $80 < x < 130$.  
Fig.  \ref{FigIonPS1} and Fig.  \ref{FigElec} are show a similar time slice to Fig. 2 in \citetalias{Murphy:2010lr}.

\begin{figure}
\centering
\includegraphics[width=\columnwidth]{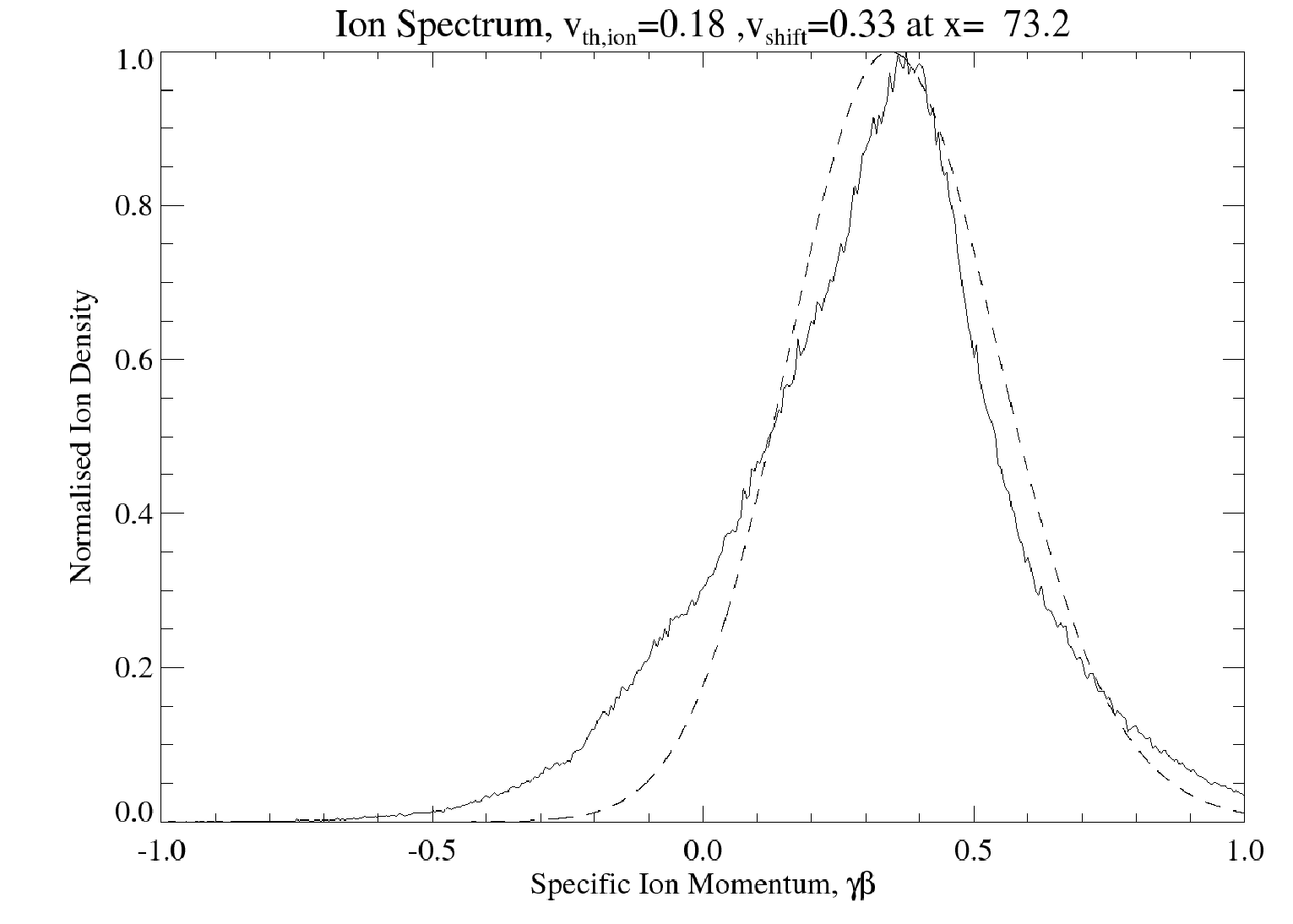}
\caption{Ion momentum distribution at $x=73.2$, normalized to its peak value, at $t=T_2$. A fit to a 
Maxwell-J\"uttner distribution with the thermal speed 0.18c and the mean speed 0.33c is overplotted.}
\label{FigJuttnerIont2}
\end{figure}

\begin{figure}
\centering
\includegraphics[width=\columnwidth]{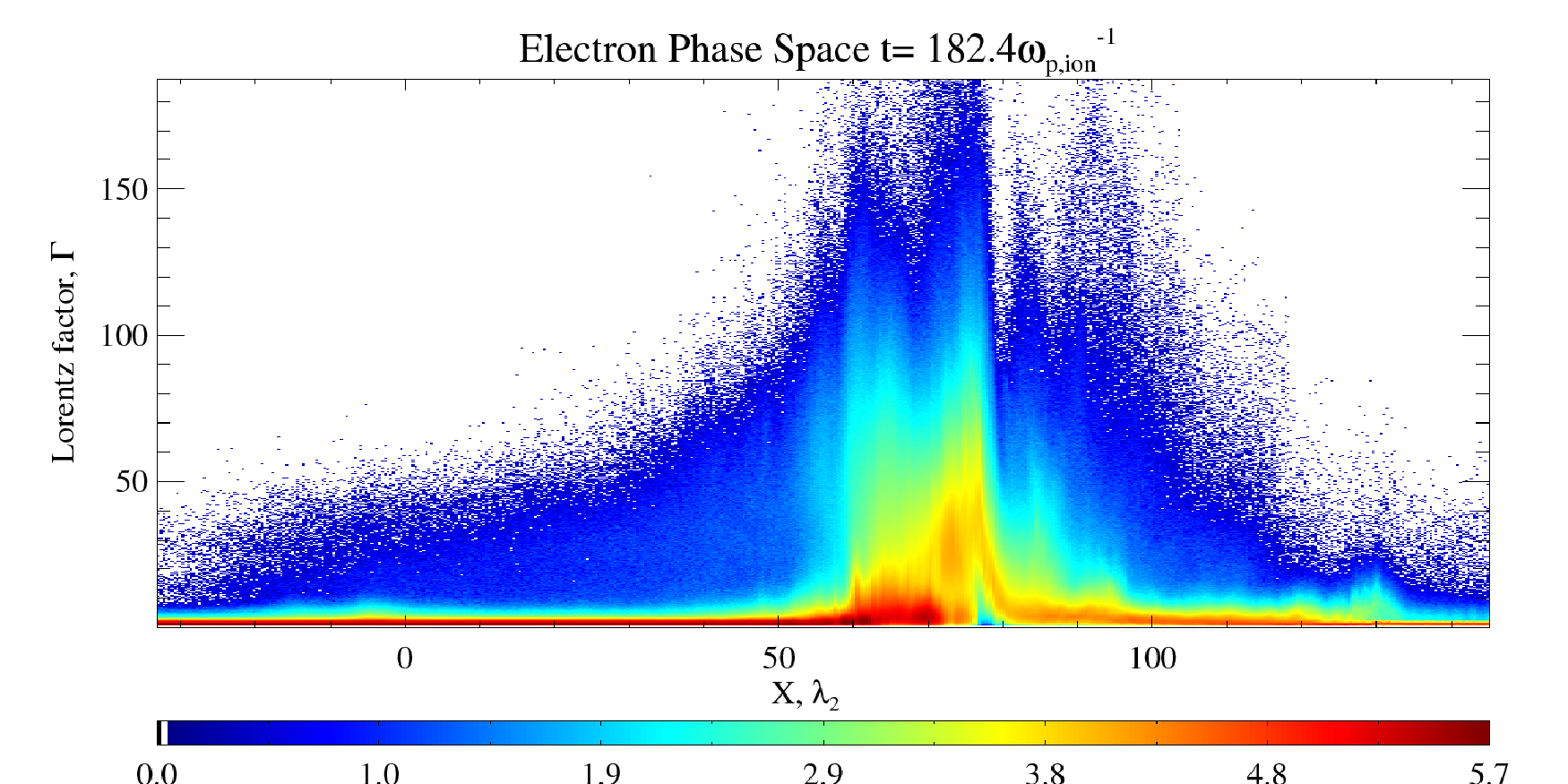}
\caption{Electron phase space distribution at $t=T_2$: Logarithm of electron density as a function of the Lorentz 
factor $\Gamma$ and x.}
\label{FigElec}
\end{figure}

First we note that the strongest interaction takes place at $x\gg 0$ and that it is tied to the front end of the
dense cloud. No energetic structure is visible at the front end of the tenuous cloud (not shown). The incoming 
plasma from the upstream, here the tenuous cloud, is reflected at $x\approx 80$, forming a shock-reflected ion 
beam. This shock reflected ion beam is hot and the reflection is not specular. A bunch of ions with values of 
$p_x$, which are comparable to the initial ones of the dense ions, is visible in the interval $100<x<120$. These 
are the dense ions, which were located ahead of the strong interaction region found at $x\approx 33$ in Fig. 
\ref{FigIonPS1}, and they have not been accelerated by it. The downstream region is characterized by a single, 
almost spatially uniform and hot ion population. The kinetic energy stored in the relative speed between the 
upstream and the downstream plasma has been converted into heat. The momentum conservation, together with the 
asymmetric cloud densities, implies that the downstream region cannot be stationary in the box frame. Indeed, 
the normalized mean speed along $x$ of the downstream ions is $\approx 0.33$ according to Fig. 
\ref{FigJuttnerIont2}, while the speed modulus of both incoming clouds is the same in the simulation frame of 
reference. The speeds of the observed forward shock that is moving to the right and the reverse shock that is 
still developing, which are given by the relative speed between the downstream plasma and the respective upstream 
plasma, must differ. 

The IIBZ to the left is characterized by the co-existence of the dense ions and the tenuous ions, which have crossed 
the front of the dense cloud prior to the formation of the shock. The mean speed modulus of the tenuous ions increases 
as we go to lower $x$ and it is close to the initial one at $x\approx -35$. This spatially varying mean momentum is a 
consequence of the shock formation. The mean speed modulus of the tenuous ions in the strong interaction region 
decreased steadily in time, as the electron acceleration became more efficient. The later in time the tenuous ions 
traversed this strong interaction region, the more energy they lost to the electrons. Once the downstream region has 
been formed, the tenuous ions can no longer cross this obstacle. The tenuous ion beam in the IIBZ is thus a transient 
effect. Eventually, a reverse shock will form between the IIBZ and the downstream, giving rise to a shock-reflected 
ion beam. 

Movie 1 shows a zoom of the time evolution of the ion distribution in $(p_x,x)$ space.
{
We note that no signal, either wave or plasma structure is detected propagating into the box either from left or right boundaries.}
The early stage shows two beams colliding.
The beams interpenetrate and then decelerate and a forward shock begins to form.
The tenuous beam is partially reflected to high velocities by the dense beam.
Left of the collision boundary, the structure is clearly heated and takes on a thermal distribution (fitted to a Maxwellian in Fig. \ref{FigJuttnerIont2}).
The lower panel shows the left moving cloud ion distribution in $(x,y)$ space.
The most significant structures at early times are the distinctive filaments, which dominate both the foreshock and the downstream region.
The filaments are sheared in the negative y direction and 
circular structures are seen to form which increase in size at the later stage of the simulation, to eventually fill the simulation box, at which point we stop the simulation.

Figure \ref{FigElec} displays the electron phase space density distribution as a function of $x$ and $\Gamma$ at $T2$.
The downstream region hosts a hot and dense electron population. The strongest electron acceleration takes place at
the boundary between the downstream region and the foreshock at $x\approx 80$. The most energetic electrons reach 
$\Gamma > 180$. Their relativistic energies are well above that of an ion with 250 electron masses and the speed $v_b$. 
The latter would have a kinetic energy, which would equal that of an electron with $\Gamma \approx 70$. A dilute cloud 
of ultrarelativistic electrons is leaking out from the downstream region into the foreshock region and into the IIBZ 
region. The distribution of the dense electrons in the IIBZ is practically unchanged and clearly separated from the 
hot population, indicating that they are not interacting strongly through beam instabilities. The turbulent tenuous 
electrons in the foreshock with $x>80$ have probably been heated by the shock-reflected ions, since they are perturbed 
in the spatial interval up to $x\approx 130$, which coincides with the cut-off of the ion beam. Some interaction may have taken place in form
of a filamentation instability between the leaking hot electrons and the incoming tenuous electrons, as it was observed
at the time T1. 

Movie 2 shows a zoom of the time evolution of the electron $(\Gamma, x)$ phase space in the upper panel, and the distribution of the left moving cloud in the lower panel.
{ We note that no plasma structure enters the box from the left or right x- boundaries.}
The acceleration to ultrarelativistic Lorentz factors occurs in a confined region close to the collision boundary.
Filamentation dominates but to a lesser extent than found in the ion clouds, due to the higher thermal velocities of the electrons in the foreshock.
At later times, the same circular structure is seen as in the ion clouds.

\begin{figure}
\centering
\includegraphics[width=\columnwidth]{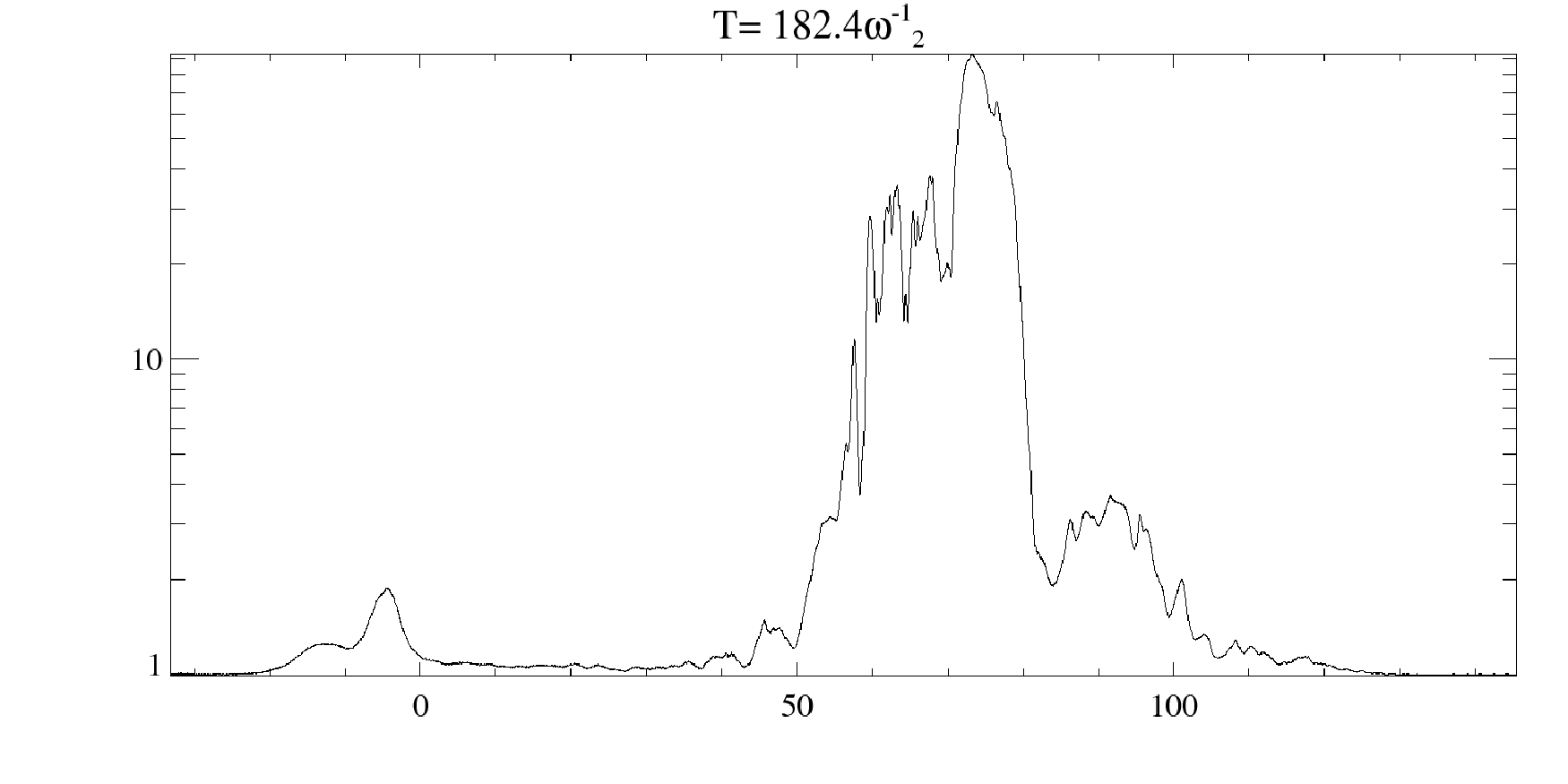}
\caption{The total magnetic energy density B$^2$ at time T2 averaged over $y$ and normalised with respect to $|{\mathbf{B}_0} |^{2}$ plotted 
against x. The field is amplified to above 50 times its initial value in a thin layer, far above the value expected 
from a shock compression.}
\label{FigFieldAmp}
\end{figure}

The magnetic energy density at the time T2 is displayed in Fig. \ref{FigFieldAmp}.
Elevated magnetic energy densities are observed in the IIBZ up to $x\approx 50$. Then the magnetic fields strengthen 
and reach a magnetic energy density plateau with $20{|\mathbf{B}_0|}^2$ at $x\approx 60$, which spans the downstream 
region up to $x\approx 70$. Then a massive peak is observed, reaching a peak value of $10^2 {|\mathbf{B}_0|}^2$ at 
$x\approx 73$. The magnetic energy density decreases rapidly as we go from this position to increasing $x$ and it 
reaches a local minimum at $x\approx 85$ with the value $\approx 2{|\mathbf{B}_0|}^2$. The second weak maximum at 
$x\approx 90$ is followed by an apparently exponentially decreasing magnetic energy density that cannot be 
distinguished from ${|\mathbf{B}_0|}^2$ at $x\approx 125$. The absolute maximum of the magnetic energy density is 
found a few ion skin depths to the left of the $x\approx 77$, in which the upstream ions are reflected in Fig. 
\ref{FigIonPST2} and where the electrons experience their strongest acceleration in Fig. \ref{FigElec}. It is thus 
the magnetic ramp in Fig. \ref{FigFieldAmp} that is responsible for the particle acceleration, but the plasma 
structure responsible for the extreme magnetic energy density is located behind it. The position $x\approx 125$, 
beyond which the magnetic field is not visibly amplified, is approximately colocated with the front of the 
shock-reflected ion beam in Fig. \ref{FigIonPST2}.
   
\begin{figure}
\centering
\includegraphics[width=\columnwidth]{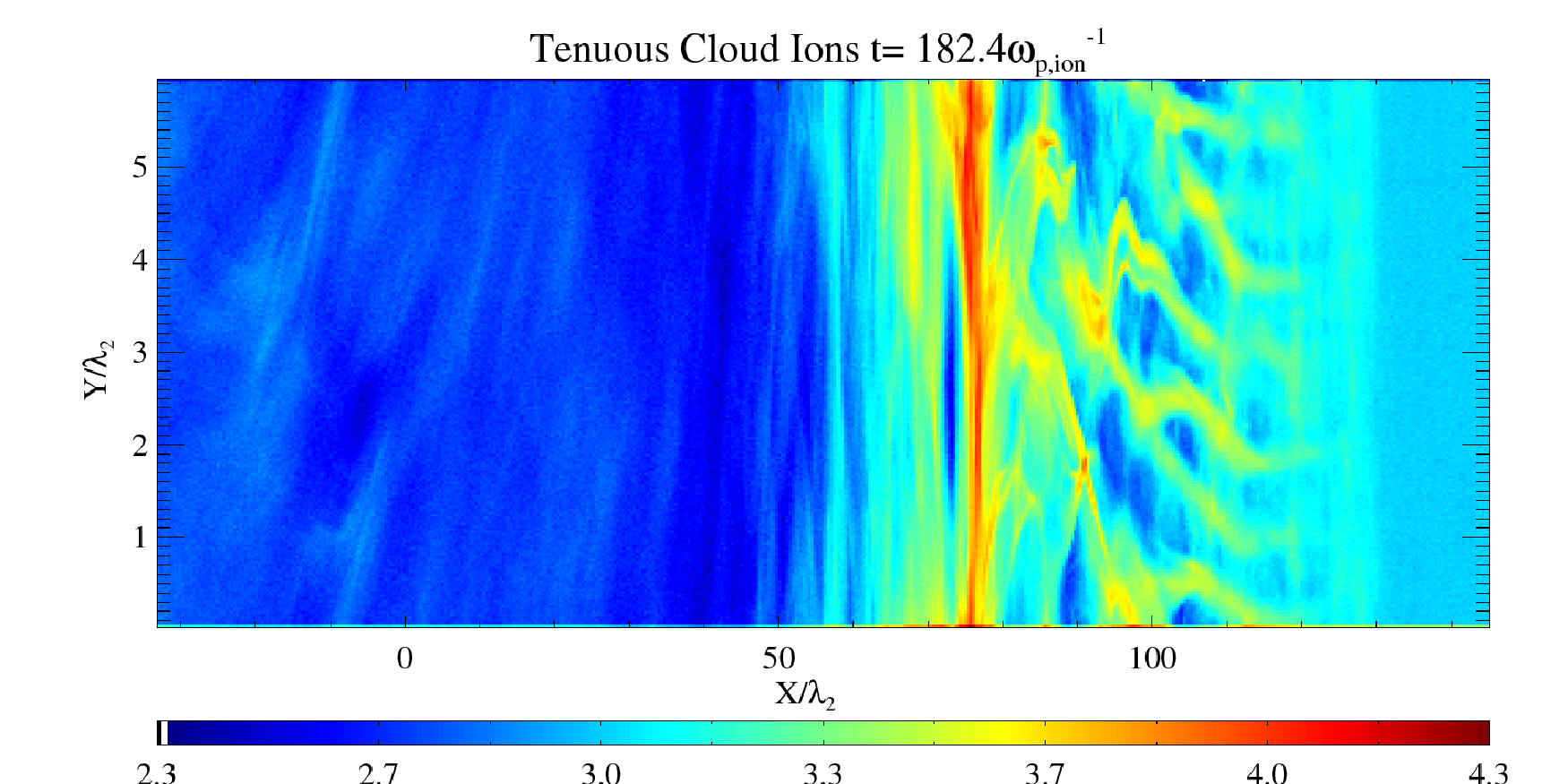}
\includegraphics[width=\columnwidth]{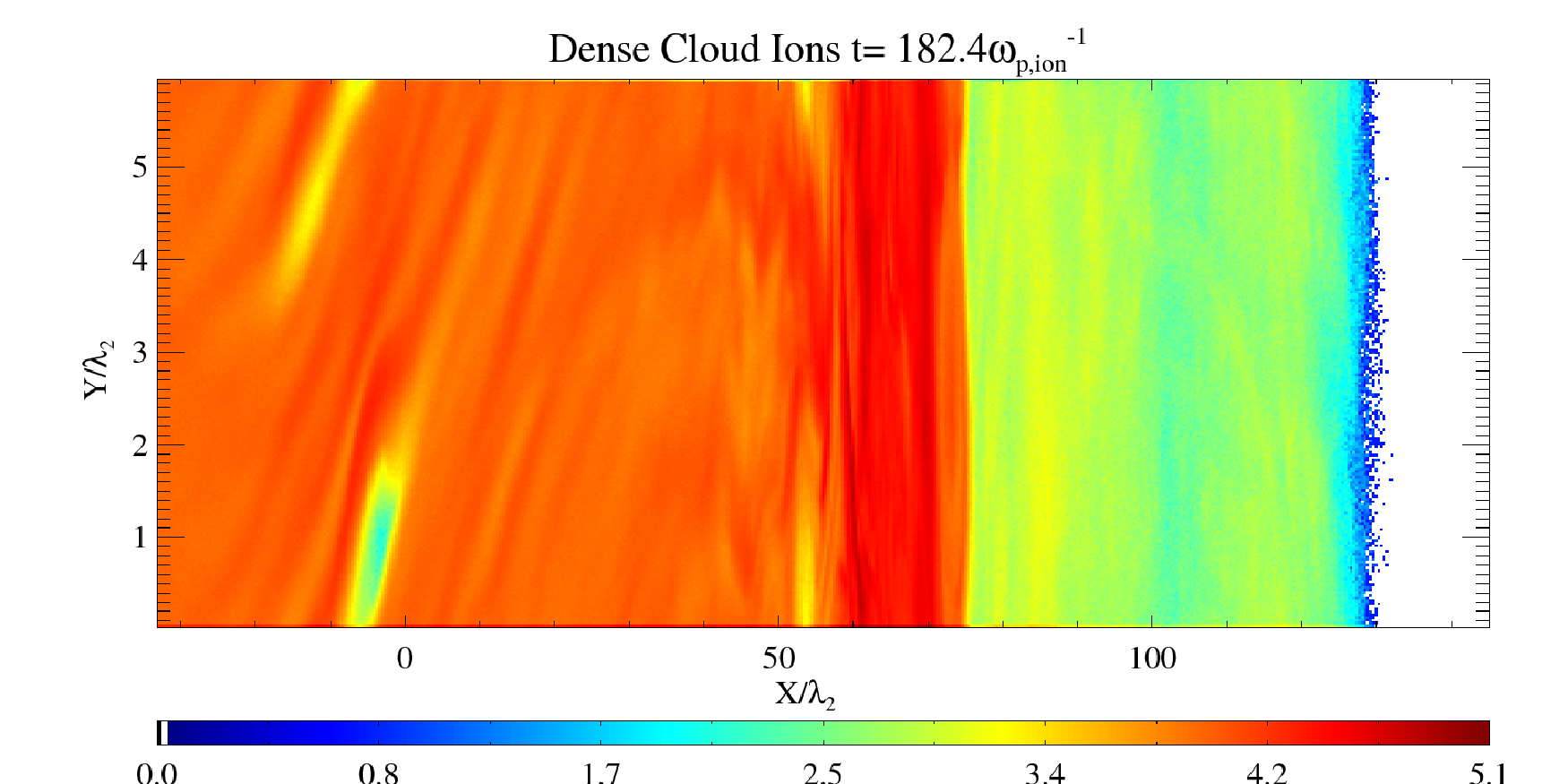}
\caption{Two dimensional logarithm of the ion densities. The upper panel 
shows the left-moving tenuous ions and the lower panel the right-moving 
dense ions at $t=T_2$.}
\label{FigIonLeftRightT2}
\end{figure}

The large magnetic energy density observed in Fig. \ref{FigFieldAmp} suggests, that the underlying currents
must be due to the ions and we analyse their density distribution in Fig. \ref{FigIonLeftRightT2}. 
A structure is observed at $-10<x<0$ and $y<2$, which is periodically wrapped around at the boundary $y=0$.
At the same position, we find a weak magnetic energy density peak in Fig. \ref{FigFieldAmp}. A strong 
density depletion is only seen in the dense ions. The movie shows that this structure can be interpreted
as the two-dimensional equivalent of a magnetic bubble. The interplay of the current filaments results in
the accumulation of magnetic energy in a localized pocket, which is convected with the dense ions. The 
pressure gradient force of this magnetic bubble expels the dense ions. The tenuous ions move with a 
relativistic speed in the rest frame of this bubble and they and their density distribution are practically 
unaffected by the magnetic pressure gradient force.

The interval $50<x<73$ with the extreme magnetic energy is characterized by the increased ion density, which 
we expect to find in the downstream region of shocks. Note that the shock compression is not unusually high. 
The ratio between the downstream ion density $\approx \exp{(5)}$ and the upstream ion density $\approx \exp{(4)}$ 
(summed over both ion species) at the forming reverse shock at $x\approx 40$ is 3 in Fig. \ref{FigIonLeftRightT2}. 
The ratio between the downstream density and the upstream ion density at the forward shock at $x\approx 75$ is 
about 4. No current filaments are observed in this interval and it is unclear from this plot what is behind this 
immense magnetic energy density. 

The tenuous ions show filamentary structures in the interval $80 < x < 130$, in which we find the weaker 
peak in the magnetic energy density in Fig. \ref{FigFieldAmp}. This peak is thus due to the ion beam 
filamentation instability. The narrow well-separated filaments, which are stretched out over tens of ion 
skin depths, can be explained with a quasi-equilibrium similar to that derived for relativistic electron 
beams by \citet{Hammer:1970}. No filaments can be found in the dense ions in this interval. The dense ions 
contribute to the fast beam, which outruns the forward shock, and this beam is too hot to react to the 
magnetic fields. The seemingly exponential growth in space of the magnetic energy density in Fig. \ref{FigFieldAmp} 
as we go from $x\approx 130$ to $x\approx 95$ may result from the combination of the exponential growth in time of 
the ion current in response of the filamentation instability with the convection to the left of the tenuous 
cloud.

\begin{figure}[t]
\centering
\includegraphics[width=\columnwidth]{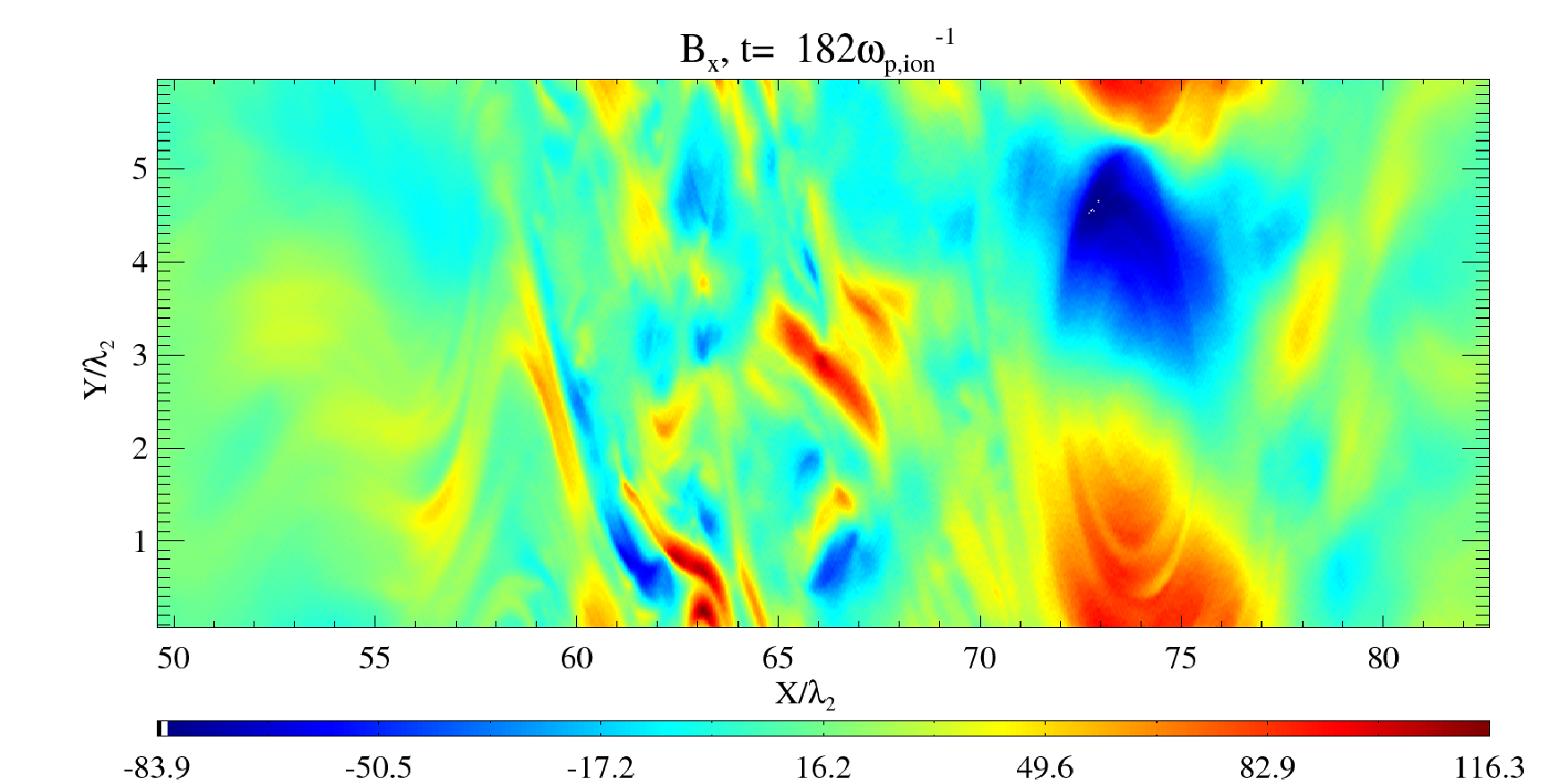}
\includegraphics[width=\columnwidth]{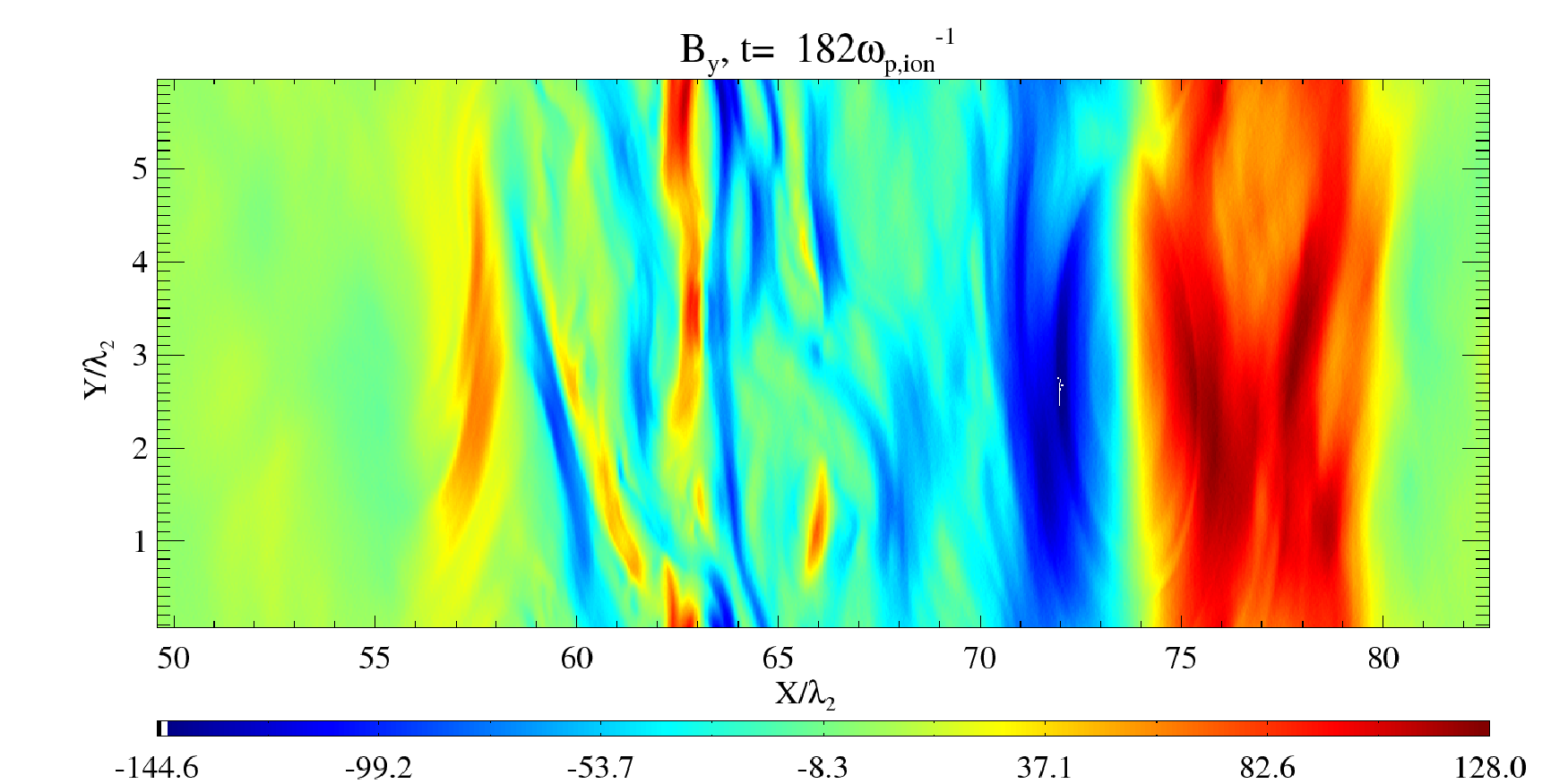}
\includegraphics[width=\columnwidth]{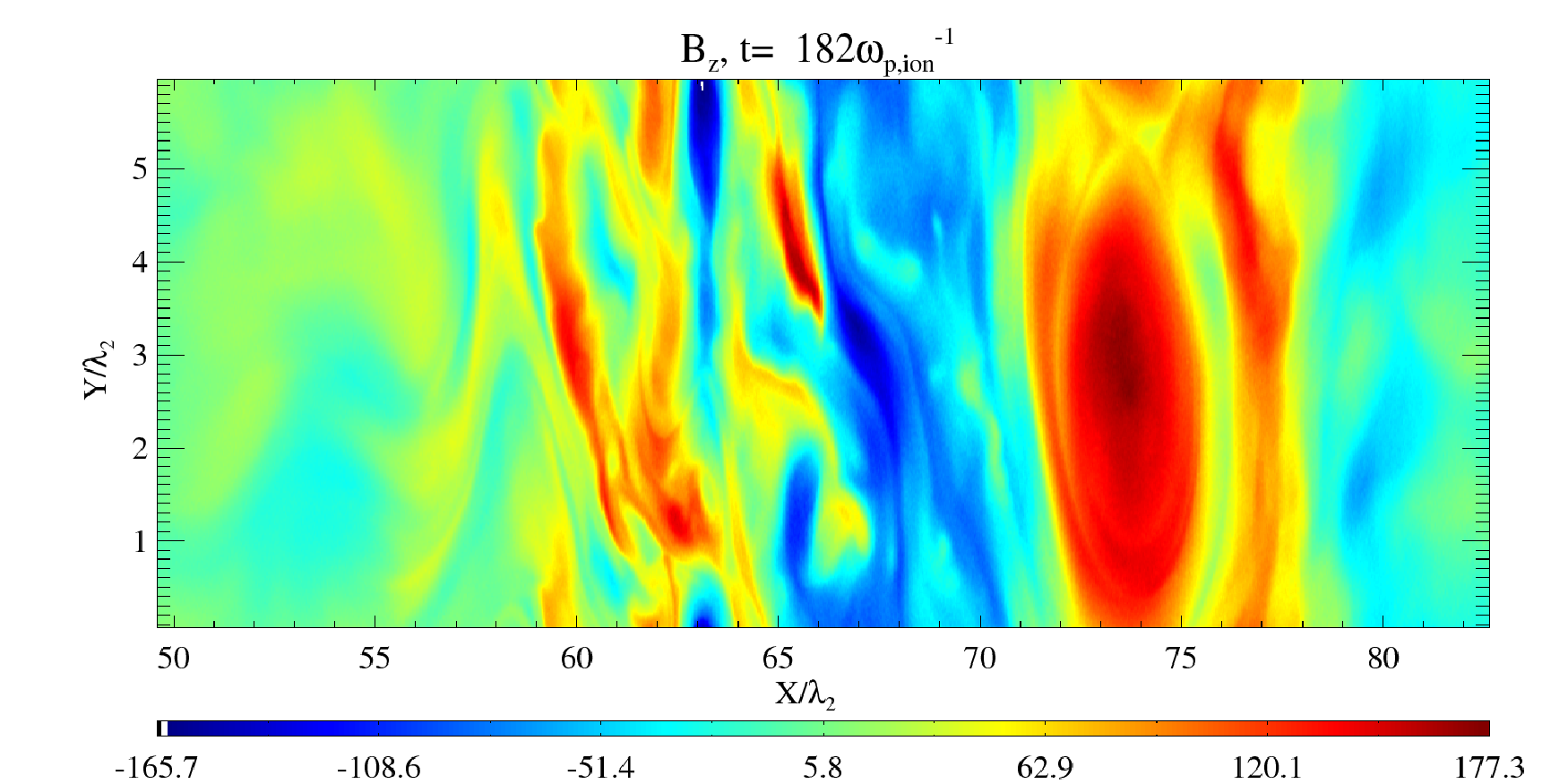}
\caption{Magnetic field components at the time T2: Zoom of a 2D linear colorscale plot of the spatial 
distribution of $B_x$ (upper panel), $B_y$ (middle panel) and $B_z$ (lower panel).}\label{FigZoomBT2}
\end{figure}

Now we turn to a more thorough examination of the structure, which is responsible for the extremely strong
magnetic field. It has previously been identified in \citetalias{Murphy:2010lr} as a flux tube. Figure \ref{FigZoomBT2} displays the three magnetic field components 
at the time T2.
The $B_x$ and the $B_y$ magnetic fields form a magnetic loop in the x-y plane for $70 < x < 77$, while 
the $B_z$ component reveals an elliptical structure within $71<x<76$. The simulation geometry implies
that we can understand this magnetic field geometry as the combination of the axial magnetic field of
a coil with an infinite extent along z and a magnetic ring that is surrounding it. Such flux tube 
distributions can be force-free, as they fullfill $\nabla \times \mathbf{B} = \alpha \mathbf{B}$ with
a constant $\alpha$. If such a structure were to form in an astrophysical jet far from any magnetic
source object, like what remains from the progenitor star, the magnetic field lines in the z-direction 
would have to be closed. A simple geometry that fullfills this necessary closure is a spheromak that 
resembles a smoke ring. 

Apart from the dominant flux tube distribution, the magnetic field shows further elongated structures 
in the interval $55 < x < 70$. The geometry of these structures is reminiscent of the end product of 
the filamentation instability, which is apparently thermalizing the downstream distribution while 
maintaining the strong magnetic fields in this domain in Fig. \ref{FigFieldAmp}. However, once the 
plasma has thermalized, these magnetic fields can no longer be upheld by plasma currents \citep{Waxman:2006qy}.

\begin{figure}[t]
\centering
\includegraphics[width=\columnwidth]{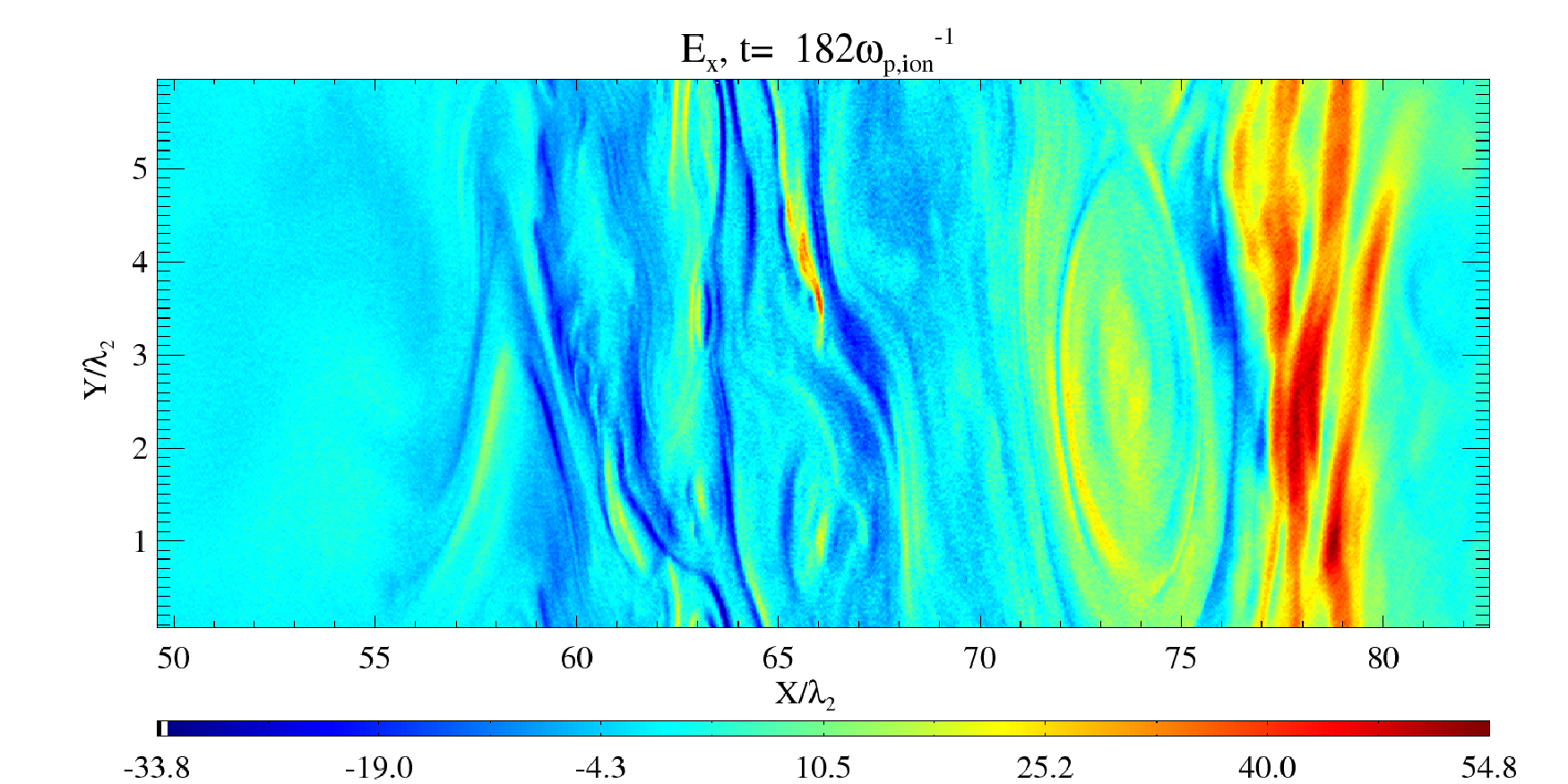}
\includegraphics[width=\columnwidth]{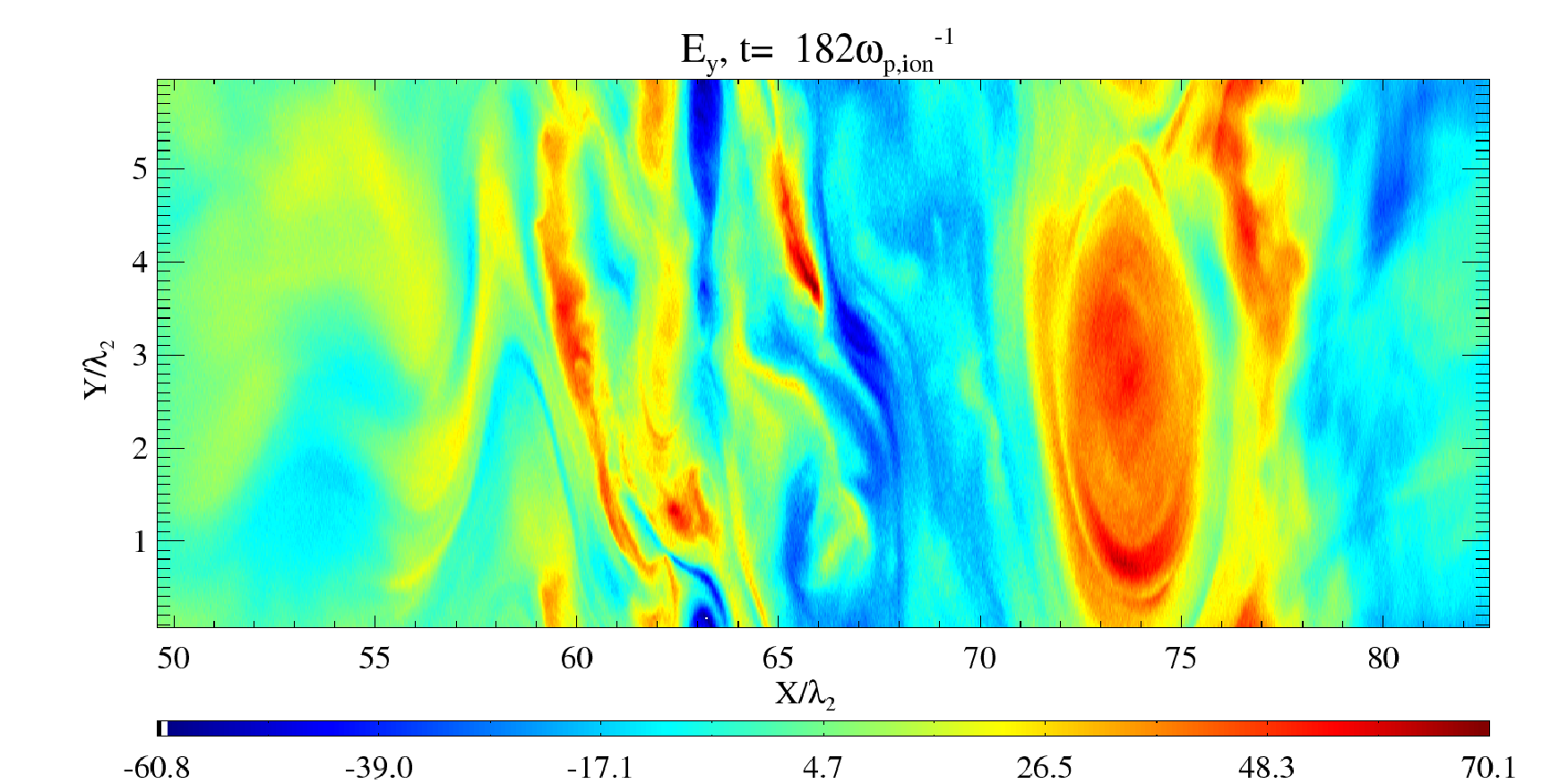}
\includegraphics[width=\columnwidth]{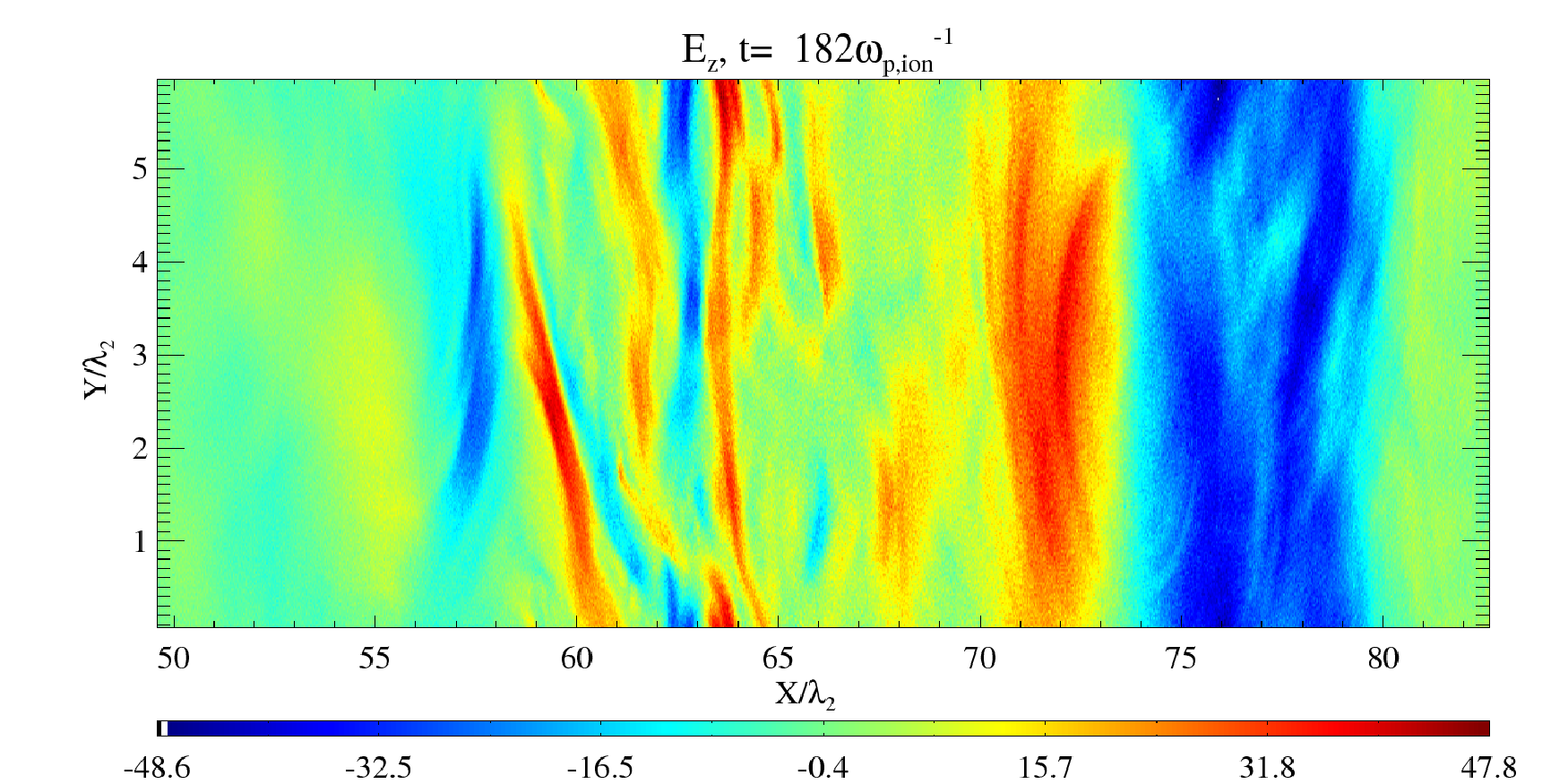}
\caption{Electric field components at the time T2: Zoom of a 2D linear colorscale plot of the spatial distribution 
of $E_x$ (upper panel), $E_y$ (middle panel) and $E_z$ (lower panel).}
\label{FigZoomET2}
\end{figure}

The strongest electron acceleration and the ion reflection by the forward shock takes place at $x\approx 77$,
where we find magnetic stripes in the $B_y$ and $B_z$ components. These structures are not closed and 
resemble the bipolar pulse we have observed at the earlier time T1. If this bipolar pulse is responsible
for the particle acceleration also at this late time, we expect to find a strong positive electric $E_x$ 
component at $x\approx 77$. Figure \ref{FigZoomET2} confirms this. While the bipolar pulse is responsible
for the electron acceleration and for the current sheet, out of which the flux tube has grown, the flux 
tube itself scatters through its magnetic field the electrons, as we can see from Fig. \ref{FigElec}.

The separation of the flux tube from the current sheet driving it may have an important consequence.
Figure \ref{FigZoomBT2} demonstrates that the periodic boundary conditions limit the growth of the flux
tube. Selecting a simulation box that is much larger in the $y$-direction may result in a larger flux 
tube, but only if the further growth is not limited by the thickness of the current sheet along
$x$. We would expect such a size limitation, if the flux tube could only exist in the current sheet. 
The current sheet would, of course, widen, if the simulation would model protons rather than the lighter 
ions. The thickness of the current sheet would, however, still be limited by the distance, over which 
the ion energy is depleted by the accelerating electrons. Here the simulation suggests that a current 
sheet ahead of the flux tube suffices to drive it. The flux tube can thus grow to a large MHD size and 
be a reservoir of magnetic energy that may not be dissipated away as quickly as that due to current 
filaments.  

The electric field furthermore demonstrates strong electric fields, which are partially correlated with 
the magnetic fields of the large flux tube and of the downstream filaments. The $E_x$ and $E_y$ components 
show, for example, the same topology as the flux tube's $B_z$ component, while the $E_z$ component resembles 
the flux tube's $B_y$ distribution.
The electric field amplitude is well below that of the magnetic field. Those of $E_x$ and $E_y$ are about half that of
their magnetic counterparts, while that of $E_z$ amounts to a third of $B_z$. The electric energy density will
thus be about 20\% of the magnetic one.

Movie 3 shows a zoom of the time evolution of all six components of the electric and magnetic fields.
{ We note that no signal, either wave or plasma structure, can be seen entering the simulation box on the left or right x-boundaries.}
In the lower four panels, $E_y, E_z$ are clearly seen to be in phase and antiphase, respectively with the $B_z, B_y$ components.
Initially a strong dipolar field dominates, and filaments are clearly visible in the $E_y, B_z$ components.
The components parallel to the shock normal  ($B_x, E_x$) are considerably more fragmented than their perpendicular counterparts.

\subsection{Generation of Mid-Infrared Synchrotron emission}

The relativistic electrons are expected to emit synchrotron emission in the presence of a magnetic field. The 
synchrotron roll-off frequency is at $\nu_{rolloff}=\omega_{c} \gamma^2 $, where $\omega_c$ is the electron 
cyclotron frequency. In the simulation we have found that for a mass ratio of $R=250$ the electrons are accelerated 
to Lorentz factors of $\sim 200$. We may assume from a comparison of the 1D PIC simulations with a mass ratio
100 and 400 in Refs. \cite{Bessho:1999sf} and \cite{Dieckmann:2008dp} that the peak electron Lorentz factor is 
proportional
to the mass ratio. In this case and for a full mass ratio many electrons would reach Lorentz factors of 1500, 
this gives $1500^2 \omega_c$. The magnetic field strength within a GRB jet is unknown and we have to resort to
a guess. For an $\omega_c=\omega_{p,electron}=10^5 $Hz this gives $\approx$ 200 GHz in the rest frame of the jet. 
There is a relativistic Doppler shift ($\delta=\sqrt{ (1+\beta) / (1-\beta) }$) from the moving frame in the jet 
into the observer's (Earth) frame. Assuming a $\beta = 0.99999$ for GRB jets, giving $\delta = $ 450, we derive a 
frequency of 100 THz in the infrared frequency range. Secondary processes such as inverse Compton 
scattering from the population of $\Gamma \sim 400 $ electrons are expected to upscatter the photons to gamma 
ray energies.

\section{Discussion}
\label{Discussion}

In this paper we have examined the collision of two plasma clouds at a shock speed of 0.9c. The aim of the study 
is to gain insight into the behaviour of mildly relativistic shocks, and the associated phenomena of electron 
acceleration, magnetic field amplification, filament formation. The shock speed is at the lower end of the 
interval proposed for internal GRB shocks \citep{Piran:1999jt}. We add a quasi-parallel guiding magnetic field to 
the simulation, in order to probe the effects of a strong magnetization expected to be present.
The mass ratio has been reduced to 250 to make the simulation more computationally tractable, while retaining the 
physical mass asymmetry which ensures a reservoir of ion energy is available to contribute to particle acceleration \citep{Amato:2006lr}. The density ratio of 10 has been chosen to emphasize the effects of asymmetry in GRB shocks.
{ This paper extends \citetalias{Murphy:2010lr} discussing in more detail the conditions that result in vortex formation also examining the electromagnetic field distribution at the shock front. While \citetalias{Murphy:2010lr}  only considered the currents, in this paper we
 focus on the particle acceleration mechanisms, not covered in the previous publication.

}

The magnetized background together with the colliding clouds ensures a jump in the convection electric field will 
be present. The initial jump in the convection electric field triggers the growth of a ramp in the magnetic field 
from which particles can be accelerated. This electromagnetic reaction to the plasma cloud collision is the first 
and most rapid response by the system. This ramp is initially planar as found by \citet{Dieckmann:2010qf}, but 
later evolves through plasma instabilities into a non-planar structure. Previous studies have shown that magnetized 
shocks have little particle acceleration unless ions are present \citep{Hoshino:1992fk,Amato:2006lr}. Equally studies 
of weakly magnetized collisions have shown little or no difference to unmagnetized plasma \citep{Nishikawa:2003eu}. 
The simultaneous presence of ions and of a strong guiding magnetic field show the rapid formation of a shock, which
accelerates electrons to highly relativistic speeds and amplifies substantially the initial magnetic field.

The results shown here - in the context of the observations of polarisation in GRBs \citep{Coburn:2003ul,Steele:2009uq} provide a compelling argument for the role of a dynamically significant primordial magnetic field from the jet. Such fields, in oblique collisions, provide a mechanism to transfer energy down the mass scales from ions to electrons, allowing electrons to increase their relativistic mass until they can be injected into the Fermi acceleration mechanism.

Based on the linear dispersion relation, we hypothesise that the filamentation instability is not suppressed in this 
regime of parameter space. The nonlinear simulation supports this hypothesis, but goes further to demonstrate a 
possible solution to the longstanding problem of filament lifetimes. Evidence for vortex formation is found, possibly 
a stable solution to offset rapid filament decay mechanisms \citep{Waxman:2006qy}. Such vortices are observed in nonrelativistic plasma flows \citep{Alexandrova:2006lr} and may thus not be unlikely structures in the more
energetic astrophysical flows. The oblique shock considered here, also not unlikely in the context of GRB jets with 
helical background fields, allows sufficient transverse motion and transverse currents for a stable vortex structure 
to form through secondary instabilities. The simulation study reveals that the vortex has an internal structure akin 
to that of the cross-section of a flux tube that can be twisted into a spheromak in a 3D geometry (see MDD).
The flux tube structure has a large inertia and is bound together by magnetic tension, possibly making it more resistant 
to dissipation on kinetic scales than the smaller-scale current filaments. During the simulation the flux tube 
continuously gains mass and magnetic field until its further growth is limited by the periodic boundary conditions 
along the shock boundary.

Comparing our work with the earlier 1D simulations by \citet{Bessho:1999sf} and the 1D and 2D simulations  \citep{Dieckmann:2008dp,Dieckmann:2010qf}, we find that the near-equipartition energy acceleration predicted 
by these authors is confirmed. \citet{Martins:2009ly} in their 2D piston simulations found that the upstream 
ions ahead of the shock were filamented, extending the region of magnetic field growth. We find that the magnetic 
field growth is exponential in the foreshock region.

\citet{Dieckmann:2010qf} showed that the structures for a lower speed simulation and smaller 
magnetic-field-to-shock-normal angle are planar and 1D. The simulations presented here show a greater departure 
from one-dimensional behaviour. The greater field angle and flow speed in our work allows more motion transverse 
to the shock plane, which apparently triggers different processes in the shock transition layer.

The results shown here carry several important implications. Firstly, the filament generated magnetic field 
can be stored in magnetic vortices. 
We consider here a magnetic field that
is relatively strong in that it yields an electron cyclotron frequency
that is comparable to the electron plasma frequency. 
Secondly, oblique shocks have good acceleration properties, increasing 
electrons to near equipartition with ions. Thirdly, primordial fields are becoming increasingly accepted in GRBs 
and will have a decisive affect on the dynamics of plasma internal shock, acceleration of electrons and their 
injection into the Fermi mechanism. Fourthly, as is becoming evident \citep{Lemoine:2010lr,Bret:2010fj} a reduced 
mass ratio has an important effect on the results of PIC simulations and this approximation needs to be validated 
by linear theory if used. More specifically in the context considered here, heavier ions can undermine the magnetic
suppression of the filamentation instability that may work for ions with a low mass.

Radiative losses can be non-negligible for shocked collisionless plasmas \citep{Fleishman:2007uq,Schlickeiser:2007kx,Schlickeiser:2008yq}. Although the PIC framework does not take into 
account radiative processes, we can infer that at least the radiative cooling by synchrotron-type emissions should 
not affect the conclusions propounded in this paper. The ions do not radiate significant energy and only a fraction 
of the electrons will convert some energy to low-energy photons through synchrotron cooling, in particular because 
the individual particles only spend a short amount of time in the high magnetic field region. The interaction
between the relativistic electrons and the photon seed that is not captured by PIC simulations may, however, result
in stronger energy losses.

\begin{acknowledgements}

The project is supported by Science Foundation Ireland grant number 08/RFP/ PHY1694, by the Swedish Vetenskapsr\aa det and by Projects
ENE2009-09276 of the Spanish Ministerio de Educaci\'{o}n y Ciencia and PAI08-0182-3162 of the Consejer\'{i}a 
de Educaci\'{o}n y Ciencia de la Junta de Comunidades de Castilla-La Mancha. The authors wish to acknowledge 
the SFI/HEA Irish Centre for High-End Computing (ICHEC) for the provision of computational facilities and support. 
The Plasma Simulation Code (PSC) was developed by Prof. Hartmut Ruhl.
\end{acknowledgements}

\bibliography{pic}

\end{document}